\documentclass[preprint,12pt, 3p]{elsarticle}

\usepackage{rotating}
\usepackage{graphicx}
\usepackage{booktabs}
\usepackage{amsmath}
\usepackage{amssymb}
\usepackage{adjustbox}
\usepackage{multirow}
\usepackage{subcaption}
\usepackage{xcolor}
\usepackage{float}
\usepackage[normalem]{ulem}
\useunder{\uline}{\ul}{}

\definecolor{lightred}{RGB}{255,153,153}
\definecolor{darkred}{RGB}{139,0,0}
\definecolor{lightgreen}{RGB}{144,238,144}
\definecolor{darkgreen}{RGB}{0,100,0}
\definecolor{lightblue}{RGB}{135,206,250}
\definecolor{darkblue}{RGB}{0,0,139}
\definecolor{brown}{RGB}{165,42,42}

\newcommand{\AS}[1]{{#1}}

\begin{document}

\begin{frontmatter}
    \title{GrIT: Group Informed Transformer for Sequential Recommendation}    

    \author[1]{Adamya Shyam}
    \ead{ashyam@cs.du.ac.in}
    
    \author[2]{Venkateswara Rao Kagita}
    \ead{venkat.kagita@nitw.ac.in}
    
    \author[1]{Bharti Rana}
    \ead{bharti@cs.du.ac.in}

    \author[1]{Vikas Kumar\corref{cor1}}
    \ead{vikas@cs.du.ac.in}
    
    \address[1]{University of Delhi, Delhi, India}
    \address[2]{National Institute of Technology, Warangal, India}
    
    \cortext[cor1]{Corresponding author}
    
    \begin{abstract}
    Sequential recommender systems aim to predict a user's future interests by extracting temporal patterns from their behavioral history. Existing approaches typically employ transformer-based architectures to process long sequences of user interactions, capturing preference shifts by modeling temporal relationships between items. 
    However, these methods often overlook the influence of group-level features that capture the collective behavior of similar users. We hypothesize that explicitly modeling temporally evolving group features alongside individual user histories can significantly enhance next-item recommendation. Our approach introduces latent group representations, where each user's affiliation to these groups is modeled through learnable, time-varying membership weights. The membership weights at each timestep are computed by modeling shifts in user preferences through their interaction history, where we incorporate both short-term and long-term user preferences. We extract a set of statistical features that capture the dynamics of user behavior and further refine them through a series of transformations to produce the final drift-aware membership weights. 
    A group-based representation is derived by weighting latent group embeddings with the learned membership scores. This representation is integrated with the user's sequential representation within the transformer block to jointly capture personal and group-level temporal dynamics, producing richer embeddings that lead to more accurate, context-aware recommendations. We validate the effectiveness of our approach through extensive experiments on five benchmark datasets, where it consistently outperforms state-of-the-art sequential recommendation methods.
    \end{abstract}
    

\end{frontmatter}

\section{Introduction}
The proliferation of information technology has resulted in an unprecedented surge of online content, making it increasingly challenging for users to identify relevant information. Recommender systems help alleviate this issue by guiding users toward personalized content, thereby enhancing user engagement and satisfaction across diverse digital platforms~\cite{lu2015recommender, ko2022survey}. Traditional approaches such as collaborative filtering and content-based recommender systems address this by capturing users' long-term preferences, often treating each item in a user's interaction history as equally important~\cite{bobadilla2013recommender}. However, these traditional approaches fail to account for the dynamic and temporal nature of user interests, which can shift rapidly due to factors such as trends, context, or recent experiences. As a result, these systems struggle to adapt to evolving user preferences, leading to suboptimal recommendation accuracy. For example, consider a user with a long-standing history of watching thriller movies, while her recent preferences align more with comedy. However, due to the sheer volume of her older interactions, the system fails to account for this shift in preference. 

Sequential recommender systems (SRSs) have become indispensable in modern recommendation scenarios due to their ability to model evolving user preferences. Unlike traditional recommendation approaches that treat user interactions as static, SRSs explicitly capture temporal dynamics in user interaction sequences, accounting for preference shifts, contextual influences, and interest transitions~\cite{fang2020deep}. This makes SRSs particularly effective for dynamic domains such as e-commerce platforms, streaming services, and digital news portals, where accurately predicting the next relevant item is crucial. To achieve this, contemporary SRSs implementations predominantly leverage neural network architectures that learn user–item representations in shared latent spaces to capture temporal dependencies, with methods evolving through several key architectural paradigms~\cite{yoon2023evolution}.

Early approaches primarily adopted recurrent neural network (RNN) to model sequential dependencies and convolution neural network (CNN) architectures to capture local interaction patterns. Donkers et al.~\cite{donkers2017sequential} propose a user-aware model based on gated recurrent unit (GRU) that integrates a static user context at each time step. This model also includes an attention mechanism to dynamically adjust focus between the user's profile and their most recent item interactions. Cui et al.~\cite{cui2018mv} address the item cold-start problem by incorporating multi-modal features through a marginalized denoising autoencoder framework. Their approach explores two architectural strategies for modeling user interests using long short-term memory (LSTM) network, one employing separate LSTM units for latent and fused vectors and another using a single jointly trained LSTM unit. Xu et al.~\cite{xu2019recurrent} propose RCNN, a hybrid architecture that combines RNN and CNN for sequential recommendation, applying horizontal and vertical convolutional filters over hidden states to capture non-linear feature interactions and local patterns. Despite their effectiveness, RNN- and CNN-based approaches struggle to capture long-range dependencies and facilitate parallel computation.

More recently, transformer-based models have been extensively applied to sequential recommendation, offering enhanced parallelization and greater flexibility in modeling long-range dependencies. Kang et al.~\cite{kang2018self} propose a self-attention-based sequential recommender (SASRec) that leverages the self-attention mechanism to selectively focus on relevant past items in a user’s interaction sequence. The attention mechanism aids in focusing on short-term transitions for sparse data, as well as on long-term dependencies in dense ones. Li et al.~\cite{li2020time} propose TiSASRec, a time-interval aware sequential model that leverages both absolute item positions and relative time intervals to learn temporal relationships between interactions. Sun et al.~\cite{sun2019bert4rec} propose BERT4Rec, a sequential recommendation model based on bidirectional encoder representations from transformers. Unlike unidirectional attention-based models, BERT4Rec employs bidirectional self-attention to capture user behavior patterns more effectively. 
To achieve efficient computation, Li et al.~\cite{li2023strec} propose STRec, a sparse transformer architecture that replaces self-attention with cross-attention by employing a novel sampling strategy to extract representative interactions that serve as queries in the cross-attention mechanism. Several other works have modified the transformer architecture to perform the sequential recommendation task~\cite{qiu2022contrastive, du2023frequency, shin2024attentive}.

Although existing neural architectures for sequential recommendation demonstrate strong performance, they often restrict the modeling of user–item sequential interactions to user-specific and item-specific features. This narrow focus overlooks the fact that in many real-world recommendation scenarios, users are influenced not only by their individual interaction histories but also by the collective preferences of the groups to which they belong. These group associations can be diverse, where users may simultaneously belong to multiple distinct groups, each representing different facets of their interests. For example, a user might align with a group favoring \textit{action} films while also sharing preferences with another group inclined toward \textit{romantic comedy}. Furthermore, these groups are inherently dynamic, evolving over time as members consume new content and adapt to emerging trends, which in turn alters a user's affinity to the group. These alterations in user affinity directly influence the group embedding, which must also be dynamic to accurately reflect shifting preferences. For example, a user initially aligned with a group favoring \textit{action} films may start watching more \textit{sci-fi} titles, as other members of the action group also explore sci-fi, the group embedding should evolve to incorporate this emerging preference. Capturing these temporally evolving group features can provide valuable contextual information beyond what is available from individual sequences alone. By estimating a user’s time-varying affinity to different groups, a group-level representation can be derived, encapsulating the shared behavioral patterns of similar users while adapting to temporal changes. We hypothesize that integrating temporally evolving group representations into sequential models enhances prediction accuracy and better captures dynamic user preferences.

To validate this hypothesis, we propose \AS{\textbf{Gr}oup-\textbf{I}nformed \textbf{T}ransformer (\textit{GrIT})}, a sequential recommendation framework 
that explicitly models temporally evolving group features alongside individual user interaction histories. The model first constructs latent group representations and learns time-varying membership weights for each user, derived from statistical features over both short- and long-term interaction windows. 
At each time step, a group-level representation is computed by aggregating latent group embeddings according to the user’s current membership weights. This group representation is then integrated with the user’s individual sequential representation within a transformer-based architecture, enabling the model to jointly capture personal and group-level temporal dynamics. The unified representation is finally used to rank candidate items for next-item prediction.

The major contributions of the proposed framework can be summarized as follows.
\begin{itemize}
    \item We propose \textit{GrIT}, a novel sequential recommendation model that captures evolving user group memberships and preference trajectories.
    \item We design a robust group membership inference method that leverages multiple statistical features to construct group-based user representations.
    \item We integrate group-aware representations with individual sequence modeling using a transformer-based encoder, enabling context-rich and adaptive recommendations.
    \item We demonstrate the effectiveness of \textit{GrIT} through extensive experiments over five benchmark datasets, showing significant improvements over baseline approaches.
\end{itemize}

The rest of the manuscript is organized as follows. We discuss the existing sequential recommendation techniques in Section \ref{relatedWork}. The proposed \textit{GrIT} approach is introduced in Section \ref{proposedGrIT}. Section \ref{expSetup} describes the comprehensive experimental setup. We present and analyze our experimental results in Section \ref{sec:rnd}. Finally, in Section \ref{concFuture}, we conclude the paper and discuss future research directions.

\section{Related Works}
\label{relatedWork}
Sequential recommendation techniques aim to predict users' future preferences by analyzing their historical interactions in chronological order. The fundamental concept behind it is that users' current preferences are strongly influenced by their recent actions. With the evolution of deep learning (DL) techniques, DL-based approaches have demonstrated superior performance as compared to traditional methods~\cite{yoon2023evolution}. The ability of DL methods to capture complex, non-linear patterns helps in modeling the user behavior sequences effectively. In this section, we discuss the existing DL-based sequential recommendation techniques. 

Zhou et al.~\cite{zhou2022filter} propose a multi-layer perceptron-based model, FMLP-Rec, with learnable filters to address the noise in logged user behavior data. The work employs the Fast Fourier Transform (FFT) for converting item representations to the frequency domain and then applies learnable filters to remove noise. The frequency domain filtering technique aids the process of distinguishing between meaningful patterns and noise in the user behavior sequences. The denoised features are then converted back using inverse FFT to provide predictions. In another work, Gao et al.~\cite{gao2024smlp4rec} introduce a Sparse MLP framework (SMLP4Rec) that employs a tri-directional information fusion mechanism to capture sequential, cross-channel, and cross-feature correlations jointly. The framework leverages the cross-feature correlations to model the relationships between heterogeneous feature types, while the cross-channel correlations aid in understanding the semantics of individual feature embeddings. Treating the sequence of items as an image in the latent space, Tang et al.~\cite{tang2018personalized} propose the Convolutional Sequence Embedding Recommendation (Caser) model to extract sequential patterns. The model incorporates horizontal filters to identify union-level sequential patterns where combinations of previous items collectively influence future preferences, and vertical filters to capture point-level patterns by learning weighted aggregations across item embeddings. This architecture enables flexible modeling of both local sequential dependencies and skip behaviors, where non-adjacent items in the sequence can still maintain predictive influence on future recommendations.
Yan et al.~\cite{yan2019cosrec} propose a 2D convolutional neural network framework for sequential recommendation, CosRec. The authors introduce a pairwise encoding module that transforms item sequences into a three-way tensor, where each element represents the concatenated embeddings of item pairs, enabling flexible interactions among non-adjacent items. The resulting tensor is then processed through a CNN to extract the high-level sequential features. 

Multiple approaches also utilize recurrent neural networks (RNN) to model the sequential data due to their ability to maintain hidden states that capture information from previous time points~\cite{liu2016context, donkers2017sequential, cui2018mv}. Along with static historical preferences, Yuan et al.~\cite{yuan2020attention} incorporate the time and location of interactions, time intervals between interactions, and correlations between items to extract sequential patterns. Their proposed attention-based context-aware model, ACA-GRU, modifies the gated recurrent unit (GRU) architecture to incorporate these four types of contextual information. Duan et al.~\cite{duan2023long} replace the gates of the long short-term memory (LSTM) architecture with attention mechanisms to capture the correlations between current items and historical behaviors. The LSTeM model enhances LSTM memory by learning a dynamic global memory embedding from user historical information. This embedding is utilized with a recover gate to retrieve information previously discarded but useful in the current context. 

\AS{In recent times, transformer-based architectures have emerged as the dominant paradigm for sequential recommendation~\cite{kang2018self, sun2019bert4rec, shin2024attentive}. To address the representation degeneration problem, Qiu et al.~\cite{qiu2022contrastive} introduce a dual contrastive learning approach, DuoRec. The model employs a contrastive regularization objective with dropout-based model-level augmentation and supervised positive sampling to improve the embedding uniformity and semantic consistency. Shin et al.~\cite{shin2024attentive} introduce BSARec, which leverages a Fourier-based attentive inductive bias to mitigate the oversmoothing problem in self-attention. The model utilizes a frequency rescaler to disentangle low- and high-frequency components and adaptively combines them. A tunable coefficient is used to balance between inductive bias and self-attention components. Several other studies have adapted transformer architectures for sequential recommendation tasks~\cite{li2020time, du2023frequency}.}

\section{Proposed Approach}
\label{proposedGrIT}
In this section, we present our proposed method, \textit{GrIT}, a novel group-informed transformer for sequential recommendation that incorporates group-derived features alongside sequence-derived features to enhance next-item prediction. Unlike conventional SRS models that rely solely on user- and item-specific embeddings, our method captures both personalized sequential patterns and collective group-level dynamics. These group dynamics are crucial for understanding how user preferences evolve over time as new items are consumed and trends emerge. To achieve this, we design a transformer-based encoder that unifies sequence modeling with group modeling within a single block. Specifically, the encoder block in \textit{GrIT} consists of four main stages. First, the input item sequence is processed by a sequence encoder, where item embeddings are augmented with learnable positional encodings and passed through a normalization and dropout layer to produce enhanced item representations. Second, we process these enhanced embeddings in parallel to capture both sequence-based and group-aware user representations. A self-attention module handles the sequence-based modeling, while a group modeling module focuses on the group-aware representations. Third, the output of the self-attention module, after residual connection and normalization, is fused with the dynamic group-level representation to jointly capture individual and collective temporal dynamics.  Finally, this unified representation is processed through a point-wise feed-forward network, which is followed by a residual connection, layer normalization, and a dropout layer. We stack 
$L$ such transformer encoder blocks, where each block integrates sequence-based features and group-based features. This layered design allows \textit{GrIT} to progressively refine user representations by capturing both short-term and long-term dependencies across different temporal scales.

In the following subsections, we formally define the problem and then describe each of these components in detail.

\subsection{Problem Formulation}
\label{probForm}
Let $\mathcal{U} = \{u_1, u_2, \dots, u_n\}$ denote the set of $n$ users and $\mathcal{V} = \{v_1, v_2, \dots, v_m\}$ denote the set of $m$ items. For each user $u \in \mathcal{U}$, the interacted items can be ordered based on the interaction time into a sequence $S^u = \{v_1^u, v_2^u, \dots, v_{l}^u\}$, where $v_i^u \in \mathcal{V}$ denotes the $i$-th interacted item of user $u$, $v_1^u$ is the oldest interaction, $v_l^u$ is the most recent interaction, and $l$ is the total number of interactions for $u$. Given $S^u$, the goal of sequential recommendation is to predict the next item $v_{l+1}^u$ the user is most likely to interact with, by modeling the interaction patterns within $S^u$. Formally, this can be expressed as:
\begin{equation}
    v_{l+1}^u = \arg\max_{v \in \mathcal{V} \setminus S^u} f(u, S^u, v),
\end{equation}
where $f(\cdot)$ is a scoring function that measures the relevance of candidate item $v$ given the user $u$ and their interaction history $S^u$ in chronological order.

\subsection{Item Sequence Encoding}

In sequential recommendation, capturing the temporal dependencies in user-item interactions is essential for modeling evolving preferences. To capture this, we encode positional information alongside item embeddings. For each item $v \in \mathcal{V}$, we maintain a global embedding vector $\mathbf{v} \in \mathbb{R}^d$. Given a user $u$'s interaction sequence $S^u$, we construct contextualized item representations that preserve temporal ordering by augmenting item embeddings with learnable positional encodings. The sum of these is then passed through a normalization layer to stabilize training and a dropout layer to mitigate overfitting. Formally, the enhanced representation of the $i$-th item in the sequence is given by:
\begin{equation}
    \mathbf{x}_i^u = \text{Dropout}(\text{LayerNorm}(\mathbf{v}_i^u + \mathbf{p}_i)),
\end{equation}
where $\mathbf{v}_i^u \in \mathbb{R}^d$ is the embedding of the $i$-th item $v_i^u$ in $S^u$, and $\mathbf{p}_i \in \mathbb{R}^d$ is the learnable positional embedding for position $i$. We denote the resulting sequence of embeddings as $S^u_x$. 

\subsection{Modeling Dynamic Group-based Representations}
The group modeling component generates dynamic group-based representations that complement user-specific sequential features. To achieve this, we first represent abstract user communities through a set of latent group embeddings. We then learn time-varying membership weights for each user, which reflect their evolving affinity toward these groups and are derived from both their interaction history and preference drift over time. The group-based representation of a user is computed through a four-step process. First, we construct a transition sequence from the input sequence to capture time-wise drift in a user's taste. Second, we learn temporal user representations by leveraging both the input and transition sequences across long- and short-term contexts using recency-weighted statistical descriptors. Third, based on these representations, we compute time-varying membership weights that reflect a user’s dynamic affinity toward the latent groups. Finally, we aggregate the group embeddings using these weights to generate the group-level representation of the user at each time step. These steps are detailed in the subsections below.

\subsubsection{Transition Sequence Construction} 

A sequence of consumed items captures a user's past interactions but does not explicitly encode how their preferences are changing over time. For example, if a user transitions from watching several \textit{action} films to \textit{sci-fi}, the transition embedding will capture this sharp shift in genre preference, information that may be diluted if we only look at the raw item sequence.  To explicitly encode this preference drift, we construct a transition sequence that represents the change in user preferences between consecutive interactions. Specifically, the transition embedding at time $i$ for user $u$ is defined as:
\begin{equation}
    \mathbf{t}_i^u = \mathbf{x}_i^u - \mathbf{x}_{i-1}^u,
\end{equation}
where $\mathbf{x}_i^u$ and $\mathbf{x}_{i-1}^u$ are the enhanced item embeddings at the $i$-th and $(i-1)$-th positions, respectively. This transition representation, together with the original sequence embeddings, is used to derive statistical features for modeling dynamic group affinities.
We denote the resulting sequence of transition embeddings as $S^u_t$. 

\subsubsection{Learning Temporal User Representations} 

To effectively capture evolving user preferences, we construct temporal user representations that jointly encode both long-term preference stability and short-term behavioral shifts. This enables our model to balance the influence of a user's historical interests with their most recent interaction patterns. For example, a user with a strong long-term affinity for \textit{classic rock} may recently begin engaging heavily with \textit{pop} music. A single representation would struggle to capture both their long-term taste and their current, temporary shift. By learning multiple representations, our approach captures a stable profile of historical tastes while simultaneously adapting to recent interaction trends. At each time point, we derive multiple representations of a user through a three-stage process: i) defining long- and short-term interaction windows over both original and transition sequences, (ii) computing recency-weighted statistical descriptors within each window, and (iii) combining these descriptors into multiple user representations for downstream group-affinity estimation.

\noindent \textit{Long- and Short-term Interaction Windows}: Effectively modeling user behavior requires distinguishing between a user’s long-standing tastes and their short-term interests. We therefore maintain two separate interaction windows, namely the \emph{complete} window and the \emph{short-term} window. The \emph{complete} window encompasses all interactions up to time step $i$, while the \emph{short-term} window contains only the most recent $w$ interactions. For user $u$ at time step $i$, the complete window for the original sequence embeddings is defined as 
$S^{u}_{x,\mathrm{c}}(i) = \{\mathbf{x}_1^u, \mathbf{x}_2^u, \dots, \mathbf{x}_i^u\}$,  
while the short-term window is  
$S^{u}_{x,\mathrm{s}}(i) = \{\mathbf{x}_{i-w+1}^u, \dots, \mathbf{x}_i^u\}$,  
where $w$ is the short-term window size.  
Similarly, for the transition embeddings, the complete window is  
$S^{u}_{t,\mathrm{c}}(i) = \{\mathbf{t}_1^u, \mathbf{t}_2^u, \mathbf{t}_3^u, \dots, \mathbf{t}_i^u\}$, where $\mathbf{t}_1^u$ is defined as a zero vector since there is no preceding interaction for the first item in the sequence and the short-term window is  
$S^{u}_{t,\mathrm{s}}(i) = \{\mathbf{t}_{i-w+1}^u, \dots, \mathbf{t}_i^u\}$. In cases where the number of preceding items is less than the short-term window size $w$, zero-padding is applied to maintain a consistent window length.  This structured representation of historical interactions serves as the foundation for computing recency-aware statistical descriptors in the next step.

\noindent \textit{Recency-weighted Statistical Descriptors}: For each interaction window $S^{u}_{\{\cdot\},\mathrm{c}}(i)$ or $S^{u}_{\{\cdot\},\mathrm{s}}(i)$, where $\{\cdot\} \in \{x, t\}$ denotes either the original sequence embeddings or the transition embeddings, we compute recency-weighted statistics using an exponentially weighted moving average (EWMA) and its corresponding variance. The EWMA captures the central tendency of recent behaviors while giving more weight to interactions closer to the current time step, thus ensuring responsiveness to preference changes. The variance measures the dispersion of interactions within the window, providing a signal for preference stability. Specifically, low variance indicates consistent interests, while high variance reflects shifting behavior. Together, these descriptors offer a compact yet informative summary of both the direction and volatility of a user’s evolving preferences. 

Let $\{\mathbf{z}_1, \mathbf{z}_2, \dots, \mathbf{z}_l\}$ denote the ordered embeddings in the given window, with $\mathbf{z}_l$ being the most recent. Given a decay parameter $\alpha \in (0,1)$, we set $\gamma = 1 - \alpha$ and assign a weight to the $j$-th element as $\lambda_j = \gamma^{l-j}$, thereby giving higher importance to recent interactions. The EWMA mean $\boldsymbol{\mu}(i)$ and its corresponding variance $\boldsymbol{\sigma}^2_i$ at time $i$ are then computed as:

$$\boldsymbol{\mu}(i) = \frac{1}{C_i} \sum_{j=1}^{i} \lambda_j \cdot \mathbf{z}_j \quad \text{and} \quad \boldsymbol{\sigma}^2(i) = \boldsymbol{\mu}^{(2)}(i) - \boldsymbol{\mu}(i) \odot \boldsymbol{\mu}(i).$$

\noindent Here, $C_i = \sum_{j=1}^{i} \lambda_j$ is the cumulative normalization term, $\boldsymbol{\mu}^{(2)}(i) = \frac{1}{C_i} \sum_{j=1}^{i} \lambda_j \, (\mathbf{z}_j \odot \mathbf{z}_j)$ is the second-order moment, and $\odot$ denotes element-wise multiplication. The variance is clamped element-wise to be at least $\varepsilon$ for numerical stability. We denote the EWMA mean and variance for the complete and short-term windows of the original sequence embeddings as $(\boldsymbol{\mu}^{u}_{x,\mathrm{c}}(i), \boldsymbol{\sigma}^{2,u}_{x,\mathrm{c}}(i))$ and $(\boldsymbol{\mu}^{u}_{x,\mathrm{s}}(i), \boldsymbol{\sigma}^{2,u}_{x,\mathrm{s}}(i))$, respectively. Similarly, for the transition embeddings, we denote them as $(\boldsymbol{\mu}^{u}_{t,\mathrm{c}}(i), \boldsymbol{\sigma}^{2,u}_{t,\mathrm{c}}(i))$ and $(\boldsymbol{\mu}^{u}_{t,\mathrm{s}}(i), \boldsymbol{\sigma}^{2,u}_{t,\mathrm{s}}(i))$. In our implementation, we employ separate decay factors $\alpha_c$ and $\alpha_w$ for the complete and short-term windows, respectively.

\noindent \textit{Constructing User Representations}: To capture the multifaceted nature of user behavior across different temporal scales, we construct four complementary representations. These representations are derived from the statistical descriptors computed over both the original and transition sequences, and from both the long-term and short-term interaction windows. Specifically, for user $u$ at time step $i$, we have two representations, $\mathbf{f}^{u}_{x,\mathrm{c}}(i)$ and $\mathbf{f}^{u}_{t,\mathrm{c}}(i)$, which correspond to the complete-window features from the original and transition embeddings, respectively.  These complete-window representations encode the user’s long-term behavioral patterns. The other two, $\mathbf{f}^{u}_{x,\mathrm{s}}(i)$ and $\mathbf{f}^{u}_{t,\mathrm{s}}(i)$, are obtained from the short-term interaction window to capture more recent preference shifts. Each representation is formed by concatenating the current embedding with its corresponding EWMA mean and variance, followed by projection through a multi-layer perceptron (MLP). Each representation is constructed as follows:

\begin{equation}
    \mathbf{f}^{u}_{x,\mathrm{c}}(i) = \mathrm{MLP}\left(\big[\,\mathbf{x}_i^u \,;\, \boldsymbol{\mu}^{u}_{x,\mathrm{c}}(i) \,;\, \boldsymbol{\sigma}^{2,u}_{x,\mathrm{c}}(i) \big] \right ),
\end{equation}
\begin{equation}
    \mathbf{f}^{u}_{t,\mathrm{c}}(i) = \mathrm{MLP}\left(\big[\,\mathbf{t}_i^u \,;\, \boldsymbol{\mu}^{u}_{t,\mathrm{c}}(i) \,;\, \boldsymbol{\sigma}^{2,u}_{t,\mathrm{c}}(i) \big]\right ),
\end{equation}
\begin{equation}
    \mathbf{f}^{u}_{x,\mathrm{s}}(i) = \mathrm{MLP}\left(\big[\,\mathbf{x}_i^u \,;\, \boldsymbol{\mu}^{u}_{x,\mathrm{s}}(i) \,;\, \boldsymbol{\sigma}^{2,u}_{x,\mathrm{s}}(i) \big]\right ),
\end{equation}
\begin{equation}
    \mathbf{f}^{u}_{t,\mathrm{s}}(i) = \mathrm{MLP}\left(\big[\,\mathbf{t}_i^u \,;\, \boldsymbol{\mu}^{u}_{t,\mathrm{s}}(i) \,;\, \boldsymbol{\sigma}^{2,u}_{t,\mathrm{s}}(i) \big]\right ),
\end{equation}
where $[\,\cdot\,;\,\cdot\,;\,\cdot\,]$ denotes vector concatenation. Together, these representations provide complementary temporal perspectives that are both temporally grounded and statistically informed, forming the basis for dynamic group membership inference.

\subsubsection{Group Affinity Modeling} 
\label{sec:groupAffine}
Given the four complementary user representations, we next estimate a user’s affinity toward a set of $\kappa$ latent groups at each time step. Our objective is to learn the group membership probability vector $\mathbf{c}^u_i \in \mathbb{R}^{\kappa}$, which represents the likelihood of user $u$ belonging to each of the $\kappa$ groups at time step $i$.  To achieve this, we exploit the complementary nature of the four temporal user representations, $\mathbf{f}^{u}_{x,\mathrm{c}}(i)$, $\mathbf{f}^{u}_{t,\mathrm{c}}(i)$, $\mathbf{f}^{u}_{x,\mathrm{s}}(i)$, and $\mathbf{f}^{u}_{t,\mathrm{s}}(i)$.  First, we aggregate features within each temporal scale by summing the original and transition-based representations:  
\begin{equation}
    \mathbf{h}^{u}_{\mathrm{c}}(i) = \mathbf{f}^{u}_{x,\mathrm{c}}(i) + \mathbf{f}^{u}_{t,\mathrm{c}}(i),
\end{equation}
\begin{equation}
    \mathbf{h}^{u}_{\mathrm{s}}(i) = \mathbf{f}^{u}_{x,\mathrm{s}}(i) + \mathbf{f}^{u}_{t,\mathrm{s}}(i).
\end{equation}

\noindent Here, $\mathbf{h}^{u}_{\mathrm{c}}(i)$ encodes long-term behavioral tendencies, while $\mathbf{h}^{u}_{\mathrm{s}}(i)$ focuses on short-term preference shifts. This aggregation leverages the complementary nature of original and transition embeddings, where original embeddings encode absolute preference states, and transition embeddings explicitly model inter-interaction dynamics. Next, we concatenate these two aggregated features and pass them through an MLP layer followed by softmax to produce group membership probabilities:
\begin{equation}
    \mathbf{c}^u_i = \mathrm{Softmax}\left( \frac{\mathrm{MLP}([\mathbf{h}^{u}_{\mathrm{c}}(i); \mathbf{h}^{u}_{\mathrm{s}}(i)])}{\tau} \right),
\end{equation}
where $\tau$ is the temperature parameter controlling the sharpness of the distribution. Lower values of $\tau$ yield more confident (peaked) assignments, whereas higher values lead to more uniform distributions.  

\subsubsection{Group-aware User Representation Learning}  
\label{sec:group_representation}
Once the time-varying group membership probabilities $\mathbf{c}^u_i$ are obtained, we use them to derive a dynamic group-aware representation for the user at time step $i$.  
Let $\mathbf{G} \in \mathbb{R}^{d \times \kappa}$ denote the learnable latent group embedding matrix, where the $g$-th column $\mathbf{G}_g$ represents the latent representations of the $g$-th group.  
The group-aware user representation at time $i$ is then computed as a weighted combination of the group embeddings, with weights given by the membership probabilities:  
\begin{equation}
    \mathbf{g}^u_i = \sum_{g=1}^{k} c^u_{i,g} \cdot \mathbf{G}_g 
    \;=\; \mathbf{G} \cdot \mathbf{c}^u_i.
\end{equation}

\noindent The dynamic representation $\mathbf{g}^u_i$ provides a compact but expressive summary of a user's preference evolution, abstracted through latent group representations.

\subsection{Fusion of Personal and Group-derived Representation}  

To create a comprehensive user profile, we fuse the dynamically derived group-based representations with the personal sequential features obtained from a self-attention module with residual connection and normalization.  This fusion enables the model to capture a more robust understanding of user taste by simultaneously exploiting the collective behavioral patterns captured by the group-aware representation and personalized interaction dynamics captured by the self-attention module. Let $\mathbf{e}^u_i$ denote the personalized sequential features for user $u$ at time step $i$. We fuse this with $\mathbf{g}^u_i$, the group-derived representation, as follows:

\begin{equation}
    \mathbf{u}^u_i = \beta \cdot \mathbf{g}^u_i + (1 - \beta) \cdot \mathbf{e}^u_i.
\end{equation}
Here, $\beta \in [0,1]$ is a hyperparameter controlling the relative importance of group-level and personal sequential features. 

The fused representation $\mathbf{u}^u_i$ is processed through a position-wise feed-forward network with a residual connection to help prevent the vanishing gradient problem and facilitate the learning of deeper feature representations. The final representation, $\mathbf{o}^u_i$, of user $u$ at time $i$ is then obtained as:
\begin{align}
\mathbf{y}^u_i &= \mathrm{GELU}(\mathbf{W}_1 \cdot \mathbf{u}^u_i + \mathbf{b}_1), \\
\mathbf{o}^u_i &= \mathrm{LayerNorm}\left( \mathrm{Dropout}\left( \mathbf{W}_2 \cdot \mathbf{y}^u_i + \mathbf{b}_2 \right) + \mathbf{u}^u_i \right),
\end{align}

\noindent where $\mathbf{W}_1$, $\mathbf{W}_2$, $\mathbf{b}_1$, and $\mathbf{b}_2$ are learnable parameters.  This fusion mechanism ensures that the final representation $\mathbf{o}^u_i$ retains the sequential context modeled by self-attention while incorporating the temporally evolving group-derived features.

\begin{figure*}[ht]
    \centering
    \includegraphics[width=\linewidth]{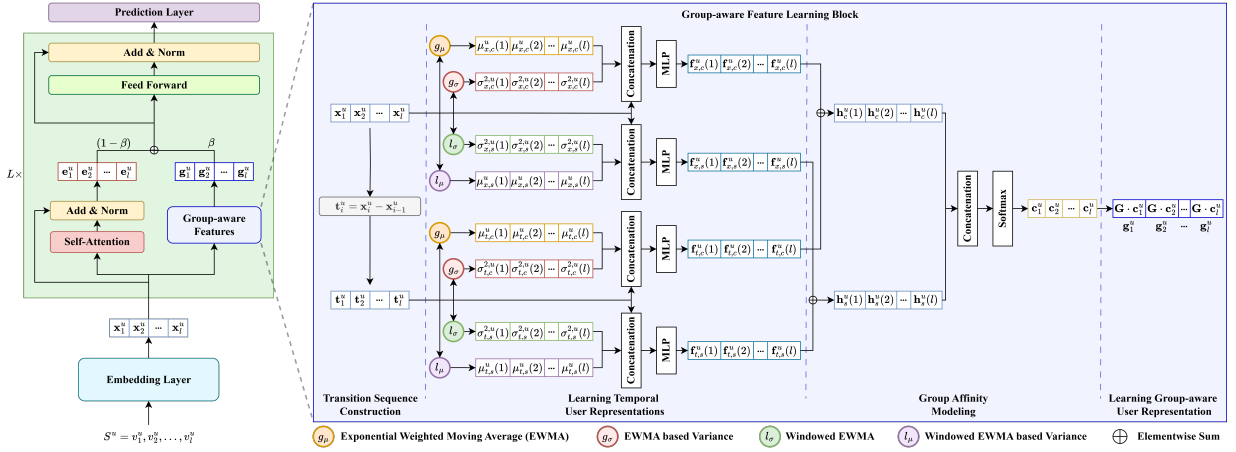}
    \caption{Overview of \textit{GrIT}.}
    \label{fig:proposedApproach}
\end{figure*}

\subsection{Prediction and Learning Objective}
\label{sec:prediction}
Given the fused user representation $\mathbf{o}^u_i$ at time step $i$, we predict the user's next action by computing a preference score for each item $v$ in the candidate set $\mathcal{V} \setminus S^u_{1:i}$.  This score is obtained by taking the dot product of the user representation with the embedding of the candidate item:

\begin{equation}
s^u_{i+1, v} = {\mathbf{o}^u_i}^\top \cdot \mathbf{v}.
\end{equation}

\noindent \textit{Learning Objective:} For training, we adopt the cross-entropy loss for the next-item prediction task. Let $\mathcal{T} = \{(u,i) \mid u\in \mathcal{U},~  v_{i+1}^u \neq \texttt{[PAD]}\}$ denote the set of valid prediction positions. We aim to minimize the difference between the predicted probabilities and the ground-truth next item $v_{i+1}^u$, as defined by the following loss function:

\begin{equation}
\mathcal{L} = -\frac{1}{|\mathcal{T}|} \sum_{(u,i) \in \mathcal{T}} \log \frac{\exp(s^u_{i+1,\,v_{i+1}^u})}{\sum_{v_j \in \mathcal{V} \setminus S^u_{1:i}} \exp(s^u_{i+1,\,v_j})}.
\end{equation}

In summary, our framework integrates personalized sequential modeling with dynamically evolving group-level behavioral cues to produce a unified, context-aware user representation. By explicitly modeling both short- and long-term interaction patterns, capturing latent group affinities, and fusing these with individual interaction histories, the proposed method enables more accurate and adaptive next-item prediction. All components are trained jointly in an end-to-end manner, ensuring seamless interaction between personal and group-derived signals. An outline of the proposed approach is presented in Figure~\ref{fig:proposedApproach}.

\section{Experimental Setup}
\label{expSetup}
This section presents the comprehensive experimental setup utilized for assessing the performance of the proposed approach against the baselines. We design a series of experiments to answer the following research questions.

\begin{itemize}
    \item \textbf{RQ1}: How does \textit{GrIT} perform compared to state-of-the-art sequential recommendation methods in terms of recommendation accuracy? (Section \ref{sec:perfCompare}) 
    \item \textbf{RQ2}: What impact do different positional encoding methods have on \textit{GrIT}’s performance? (Section \ref{sec:ablationPosEncode})
    \item \textbf{RQ3}: What is the effect of different feature types on the quality of learned group representations? (Section \ref{sec:ablationFeatImpact})
    \item \textbf{RQ4}: How distinct and well-separated are the learned group representations? (Section \ref{sec:ablationGroupDist})
    \item \textbf{RQ5}: Does user affinity to groups change over time? (Section \ref{sec:ablationGroupEvolve})
    \item \textbf{RQ6}: How sensitive is \textit{GrIT} to variations in hyperparameters? (Section \ref{sec:hyperSensitivity})
    \item \textbf{RQ7}: Are the performance differences statistically significant across datasets and baselines? (Section \ref{sec:hypoTest})
\end{itemize}

For evaluation, we employ three standard evaluation metrics, namely $Recall$, normalized discounted cumulative gain ($NDCG$), and mean reciprocal rank ($MRR$). Each metric produces values in the range $[0,1]$, with a score of $1$ indicating the best possible recommendation performance. For each evaluation metric, we report results at cutoff $k \in \{5, 10, 20\}$. For fair and comprehensive comparison, we utilize the full item set to analyze the ranking results \cite{shin2024attentive}.

\subsection{Datasets}
We evaluate the performance of our proposed approach and baseline algorithms using five publicly available datasets from Amazon\footnote{https://amazon-reviews-2023.github.io/} and GroupLens\footnote{https://grouplens.org/datasets/} repositories. Specifically, we use the  Video Games, Industrial \& Scientific, and CDs \& Vinyl datasets from Amazon, and the MovieLens 100K and MovieLens 1M datasets from GroupLens. The statistics of these datasets are presented in Table \ref{tab:datastats}.

\begin{table}[ht]
    \centering
    \renewcommand{\arraystretch}{1}
    \caption{Statistics of the considered datasets.}
    \adjustbox{max width=\linewidth}{
    \begin{tabular}{lcccc}
        \toprule
        \textbf{Dataset} & \textbf{\# Users} & \textbf{\# Items} & \textbf{\# Interactions} & \textbf{Sparsity (\%)} \\
        \midrule
        \textbf{ML-100K} & 943 & 1349 & 99287 & 92.19 \\
        \textbf{ML-1M} & 6040 & 3416 & 999611 & 95.16 \\
        \textbf{Video   Games} & 94762 & 25612 & 814586 & 99.96 \\
        \textbf{Industrial   \& Scientific} & 50985 & 25848 & 412947 & 99.96 \\
        \textbf{CDs \& Vinyl} & 123876 & 89370 & 1552764 & 99.98 \\
        \bottomrule
    \end{tabular}
    }
    \label{tab:datastats}
\end{table}

\noindent We implement 5-core filtering to reduce data sparsity and improve the robustness of learned representations. Following \cite{yue2024linear}, we handle variable-length interaction sequences by adopting a fixed sequence length $L$. Sequences longer than $L$ are divided into chunks, while shorter sequences are left-padded with a special \texttt{[mask]} token to ensure a consistent input length. We consider $L = 50$ for our experiments, consistent with the majority of baseline methods. For dataset partitioning, we employ the standard leave-one-out strategy. For each user, we chronologically order their interactions and designate the most recent interaction as the test instance and the immediately preceding interaction for validation. During the testing phase, both training and validation interactions are provided as input to the model.

\begin{table*}[!h]
    \renewcommand{\arraystretch}{1}
    \centering
    \caption{Results of the comparing algorithms $(mean ~rank)$ in terms of $Recall@k$ and $NDCG@k$. The result of the best baseline is underlined, and the Improv. (\%) denotes the relative improvement of \textit{GrIT} over it. }
    \adjustbox{width=\linewidth}{
    \begin{tabular}{llccccccccccccc}
        \hline
        \textbf{Dataset} & \textbf{Metric} & \textbf{FMLPRec} \cite{zhou2022filter} & \multicolumn{1}{l}{\textbf{}} & \textbf{DuoRec} \cite{qiu2022contrastive} & \multicolumn{1}{l}{\textbf{}} & \textbf{LinRec} \cite{liu2023linrec} & \multicolumn{1}{l}{\textbf{}} & \textbf{BSARec} \cite{shin2024attentive} & \multicolumn{1}{l}{\textbf{}} & \textbf{LRURec} \cite{yue2024linear} & \multicolumn{1}{l}{\textbf{}} & \textbf{\textit{GrIT}} & \multicolumn{1}{l}{\textbf{}} & \multicolumn{1}{l}{\textbf{Improv. (\%)}} \\ \hline
        \multirow{6}{*}{\textbf{ML-100K}} & \textbf{Recall@5} & 0.0431 & 5 & 0.0418 & 6 & 0.1052 & 4 & {\ul 0.1207} & 2 & 0.1138 & 3 & \textbf{0.1280} & 1 & 5.6387 \\
         & \textbf{NDCG@5} & 0.0295 & 5 & 0.0267 & 6 & 0.0653 & 4 & {\ul 0.0788} & 2 & 0.0764 & 3 & \textbf{0.0850} & 1 & 7.2504 \\
         & \textbf{Recall@10} & 0.0587 & 6 & 0.0700 & 5 & 0.1758 & 4 & {\ul 0.2031} & 2 & 0.2024 & 3 & \textbf{0.2099} & 1 & 3.2227 \\
         & \textbf{NDCG@10} & 0.0345 & 6 & 0.0355 & 5 & 0.0881 & 4 & 0.1050 & 3 & {\ul 0.1057} & 2 & \textbf{0.1114} & 1 & 5.1277 \\
         & \textbf{Recall@20} & 0.0917 & 6 & 0.1199 & 5 & 0.2765 & 4 & {\ul \textbf{0.3020}} & 1 & 0.2869 & 3 & 0.2957 & 2 & -2.1095 \\
         & \textbf{NDCG@20} & 0.0428 & 6 & 0.0480 & 5 & 0.1130 & 4 & {\ul 0.1298} & 2 & 0.1272 & 3 & \textbf{0.1329} & 1 & 2.2711 \\ \hline
        \multirow{6}{*}{\textbf{ML-1M}} & \textbf{Recall@5} & 0.0898 & 6 & 0.1913 & 4 & 0.1443 & 5 & 0.2185 & 3 & {\ul 0.2268} & 2 & \textbf{0.2347} & 1 & 3.3544 \\
         & \textbf{NDCG@5} & 0.0586 & 6 & 0.1328 & 4 & 0.0953 & 5 & 0.1514 & 3 & {\ul 0.1581} & 2 & \textbf{0.1651} & 1 & 4.2525 \\
         & \textbf{Recall@10} & 0.1449 & 6 & 0.2784 & 4 & 0.2223 & 5 & 0.3077 & 3 & {\ul 0.3134} & 2 & \textbf{0.3232} & 1 & 3.0215 \\
         & \textbf{NDCG@10} & 0.0763 & 6 & 0.1608 & 4 & 0.1204 & 5 & 0.1800 & 3 & {\ul 0.1860} & 2 & \textbf{0.1938} & 1 & 4.0124 \\
         & \textbf{Recall@20} & 0.2146 & 6 & 0.3792 & 4 & 0.3251 & 5 & 0.4113 & 3 & {\ul 0.4129} & 2 & \textbf{0.4248} & 1 & 2.7992 \\
         & \textbf{NDCG@20} & 0.0937 & 6 & 0.1861 & 4 & 0.1463 & 5 & 0.2061 & 3 & {\ul 0.2110} & 2 & \textbf{0.2195} & 1 & 3.8526 \\ \hline
        \multirow{6}{*}{\textbf{Industrial \&   Scientific}} & \textbf{Recall@5} & 0.0146 & 6 & 0.0317 & 3 & 0.0234 & 5 & {\ul 0.0320} & 2 & 0.0305 & 4 & \textbf{0.0337} & 1 & 5.0431 \\
         & \textbf{NDCG@5} & 0.0097 & 6 & 0.0213 & 3 & 0.0163 & 5 & {\ul 0.0221} & 2 & 0.0211 & 4 & \textbf{0.0239} & 1 & 7.6057 \\
         & \textbf{Recall@10} & 0.0215 & 6 & {\ul 0.0476} & 2 & 0.0358 & 5 & 0.0461 & 3 & 0.0448 & 4 & \textbf{0.0482} & 1 & 1.1209 \\
         & \textbf{NDCG@10} & 0.0119 & 6 & 0.0265 & 3 & 0.0203 & 5 & {\ul 0.0267} & 2 & 0.0257 & 4 & \textbf{0.0286} & 1 & 6.7324 \\
         & \textbf{Recall@20} & 0.0304 & 6 & {\ul \textbf{0.0694}} & 1 & 0.0523 & 5 & 0.0663 & 3 & 0.0647 & 4 & 0.0668 & 2 & -3.9731 \\
         & \textbf{NDCG@20} & 0.0142 & 6 & {\ul 0.0320} & 2 & 0.0244 & 5 & 0.0318 & 3 & 0.0307 & 4 & \textbf{0.0333} & 1 & 3.9442 \\ \hline
        \multirow{6}{*}{\textbf{Video Games}} & \textbf{Recall@5} & 0.0349 & 6 & 0.0697 & 3 & 0.0481 & 5 & 0.0684 & 4 & {\ul 0.0700} & 2 & \textbf{0.0709} & 1 & 1.3804 \\
         & \textbf{NDCG@5} & 0.0228 & 6 & 0.0472 & 3 & 0.0323 & 5 & 0.0467 & 4 & {\ul 0.0474} & 2 & \textbf{0.0482} & 1 & 1.6148 \\
         & \textbf{Recall@10} & 0.0541 & 6 & {\ul 0.1042} & 2 & 0.0725 & 5 & 0.0991 & 4 & 0.1036 & 3 & \textbf{0.1047} & 1 & 0.5326 \\
         & \textbf{NDCG@10} & 0.029 & 6 & {\ul 0.0584} & 2 & 0.0401 & 5 & 0.0566 & 4 & 0.0583 & 3 & \textbf{0.0588} & 1 & 0.7447 \\
         & \textbf{Recall@20} & 0.0808 & 6 & {\ul 0.1484} & 2 & 0.1066 & 5 & 0.1402 & 4 & 0.1474 & 3 & \textbf{0.1485} & 1 & 0.0071 \\
         & \textbf{NDCG@20} & 0.0357 & 6 & {\ul 0.0695} & 2 & 0.0486 & 5 & 0.0669 & 4 & 0.0693 & 3 & \textbf{0.0700} & 1 & 0.7728 \\ \hline
        \multirow{6}{*}{\textbf{CDs \& Vinyl}} & \textbf{Recall@5} & 0.0292 & 6 & 0.0651 & 4 & 0.0441 & 5 & {\ul 0.0726} & 2 & 0.0723 & 3 & \textbf{0.0730} & 1 & 0.6590 \\
         & \textbf{NDCG@5} & 0.0203 & 6 & 0.0451 & 4 & 0.0308 & 5 & {\ul 0.0514} & 2 & 0.0508 & 3 & \textbf{0.0517} & 1 & 0.5159 \\
         & \textbf{Recall@10} & 0.0398 & 6 & 0.0925 & 4 & 0.0609 & 5 & 0.0988 & 3 & {\ul 0.1001} & 2 & \textbf{0.1013} & 1 & 1.1659 \\
         & \textbf{NDCG@10} & 0.0237 & 6 & 0.0539 & 4 & 0.0362 & 5 & {\ul 0.0599} & 2 & 0.0598 & 3 & \textbf{0.0608} & 1 & 1.4825 \\
         & \textbf{Recall@20} & 0.0530 & 6 & 0.1254 & 4 & 0.0826 & 5 & 0.1302 & 3 & {\ul 0.1337} & 2 & \textbf{0.1357} & 1 & 1.4401 \\
         & \textbf{NDCG@20} & 0.0270 & 6 & 0.0622 & 4 & 0.0416 & 5 & 0.0678 & 3 & {\ul 0.0683} & 2 & \textbf{0.0695} & 1 & 2.4034 \\ \hline
    \end{tabular}
    }

    \label{tab:results}
\end{table*}

\subsection{Baselines}
To showcase the superiority of the proposed approach, we compare it with five state-of-the-art methods. The baseline methods are briefly described below.
\begin{itemize}
    \item FMLPRec~\cite{zhou2022filter}: An all-MLP sequential recommendation model that replaces self-attention with frequency-domain filtering. The model employs the Fast Fourier Transform (FFT) to convert item sequences into the frequency domain. It then uses learnable filters to reduce noise, reconstructs sequences via inverse FFT, and uses residual and feed-forward layers to capture both short- and long-term dependencies.  
    
    \item DuoRec~\cite{qiu2022contrastive}: 
    A dual contrastive learning approach that addresses representation degeneration by introducing a contrastive regularization objective. It employs dropout-based augmentation to form positive pairs and leverages supervised positives by treating sequences with the same next item as similar.  
    
    \item LinRec~\cite{liu2023linrec}: 
    A linear attention mechanism for sequential recommendation that reorders dot-product computation and applies normalization before attention calculation. It reduces complexity from quadratic to linear while effectively modeling sequential dependencies.

    \item BSARec~\cite{shin2024attentive}: 
    BSARec introduces Fourier-based attentive bias to alleviate oversmoothing in self-attention. It balances low-frequency (long-term) and high-frequency (short-term) components through a learnable frequency rescaler and integrates them with self-attention.  
    
    \item LRURec~\cite{yue2024linear}:   This method proposes linear recurrent units (LRU) by reformulating the recurrence mechanism to enable efficient parallel training and stable long-range dependency modeling. Each LRU block integrates residual connections, normalization, and position-wise feed-forward networks, similar to transformer architectures.

\end{itemize}

\subsection{Parameter Configuration}
We conduct extensive hyperparameter tuning for all comparing algorithms to ensure a fair comparison. The embedding dimension $d$ is fixed to $64$ across all methods. For \textit{GrIT}, the temperature parameter $\tau$ is set to 2, while the decay parameter $\alpha$ is set to $0.01$ for complete sequences and $0.05$ for windowed sequences, and the short-term window size is fixed at $5$. The number of groups $\kappa$ is searched in $[64, 128, 256]$, and the control parameter $\beta$ is searched in $[0.1, 0.3, 0.5, 0.7, 0.9]$. Our model employs two layers and four attention heads, while for the baselines, we follow the layer and head configurations reported in their respective papers. For all comparing algorithms, the dropout rate was searched in $[0.1, 0.2, 0.3, 0.4, 0.5]$. Hyperparameters not specified explicitly were tuned within the ranges suggested in the original implementations.

For training, we employed the \textit{AdamW} optimizer with \textit{weight decay} = 0.01, \textit{learning rate}
= 0.001, and \textit{batch size} = 256. Each model trains for up to $500$ epochs, with early stopping applied on the validation set using $Recall@10$ as the monitoring metric and a patience of $10$ to prevent overfitting. The optimal hyperparameter configuration for each model is determined by averaging $Recall@10$ and $MRR@10$ on the validation set, and the configuration yielding the highest combined score is selected as the best-performing set.

\begin{figure*}[h]
    \centering
    \includegraphics[width=\linewidth]{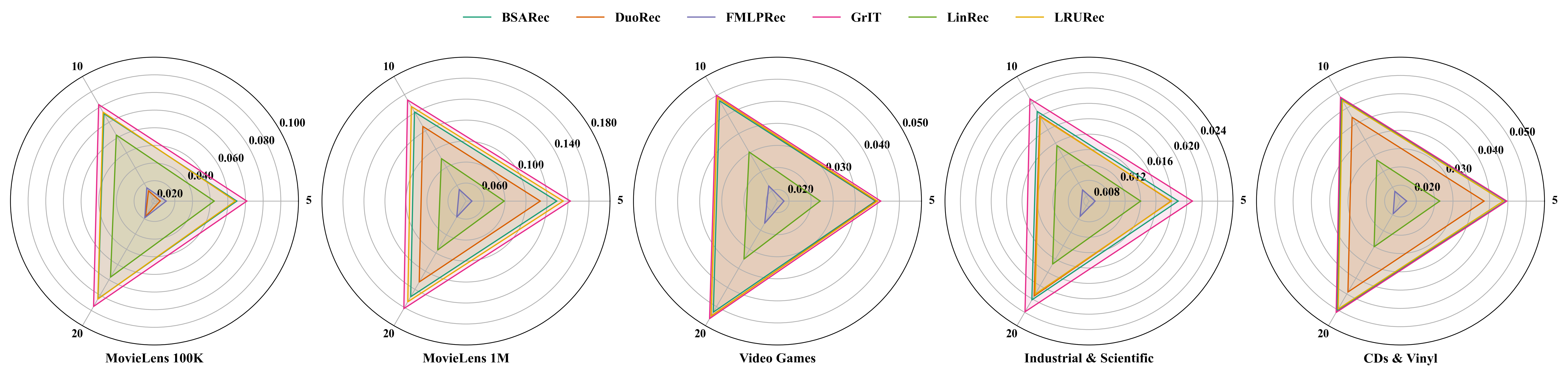}
    \caption{Performance of comparing algorithms in terms of $MRR@k$.}
    \label{fig:mrr@k}
\end{figure*}

\section{Results and Discussion}
\label{sec:rnd}
This section presents the experimental results and discusses the findings with respect to each research question.

\subsection{Overall Performance} \label{sec:perfCompare}
Table \ref{tab:results} presents a comprehensive evaluation of \textit{GrIT} compared to recent state-of-the-art baselines. \textit{GrIT} consistently outperforms the competing algorithms across all datasets in terms of $Recall@k$ and $NDCG@k$. The model demonstrates strong performance across both smaller values of $k$, capturing users' immediate preferences, and larger values of $k$, reflecting its ability to preserve long-term dependencies. Figure \ref{fig:mrr@k} further complements the results, highlighting the superior ranking quality achieved by \textit{GrIT}. Higher $MRR$ values indicate that the model is responsive to both immediate and recurring user interests. This property is particularly important for sequential recommendation scenarios where user satisfaction depends not only on accuracy but also on the timeliness and order of recommendations. The significant improvements across all three evaluation metrics underscore the efficacy of \textit{GrIT}. 

The results demonstrate that by explicitly modeling the temporal evolution of group profiles and integrating them into the recommendation process, \textit{GrIT} delivers recommendations that are simultaneously accurate and relevant. These results validate the ability of \textit{GrIT} to balance short-term responsiveness with long-term stability, establishing it as a robust solution for sequential recommendations.

\subsection{Impact of Positional Encoding Strategies} \label{sec:ablationPosEncode}
This subsection investigates how different positional encoding strategies influence the performance of \textit{GrIT} in capturing sequential dependencies. Since user interactions are inherently ordered, positional encodings are essential for enabling the model to distinguish between items occurring at different timesteps. Accordingly,  we evaluate three strategies within the \textit{GrIT} framework: (a) Fixed Sinusoidal Positional Encoding ($\mathcal{P}_{s}$), (b) Learnable Sinusoidal Positional Encoding ($\mathcal{P}_{ls}$), and (c) Fully Learnable Positional Encoding ($\mathcal{P}_{fl}$).

\begin{figure}[h]
    \centering
    \includegraphics[width=\linewidth]{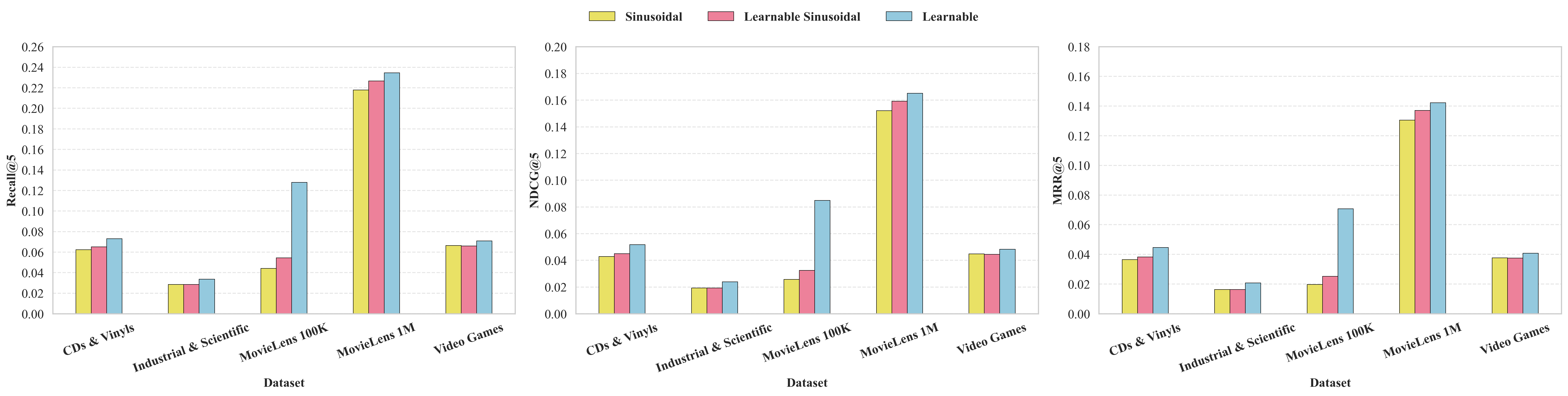}
    \caption{Influence of positional encoding strategies.}
    \label{fig:posEncode}
\end{figure}

The results presented in Figure \ref{fig:posEncode} reveal a clear performance hierarchy among these approaches. \textit{GrIT} achieves the best results with $\mathcal{P}_{fl}$, followed by $\mathcal{P}_{ls}$, and finally with $\mathcal{P}_{s}$. The performance of \textit{GrIT} significantly drops with $\mathcal{P}_{s}$, highlighting that learnable encodings allow the model to adapt positional information to dataset-specific interaction dynamics, rather than relying on predefined patterns solely. Overall, the superiority of learnable encodings demonstrates that user interaction sequences benefit from flexible representations of temporal order, effectively capturing both local and global sequential dependencies.

\subsection{Impact of Feature Combinations on Group Representations}
\label{sec:ablationFeatImpact}
To analyze the contribution of different statistical features in group representation learning, we conduct an ablation study by incrementally incorporating these features into the group representation module. Specifically, the four feature categories are (i) complete sequence features, (ii) complete transition features, (iii) short-term window sequence features, and (iv) short-term transition features. We begin by evaluating a vanilla $GrIT$ that uses no group features, followed by versions that utilize each feature type individually. Subsequently, we construct combinations of the best-performing single feature with others, moving from pairs to triplets, and finally incorporating all four features together.

\begin{figure}[h]
    \centering
    \includegraphics[width=\linewidth]{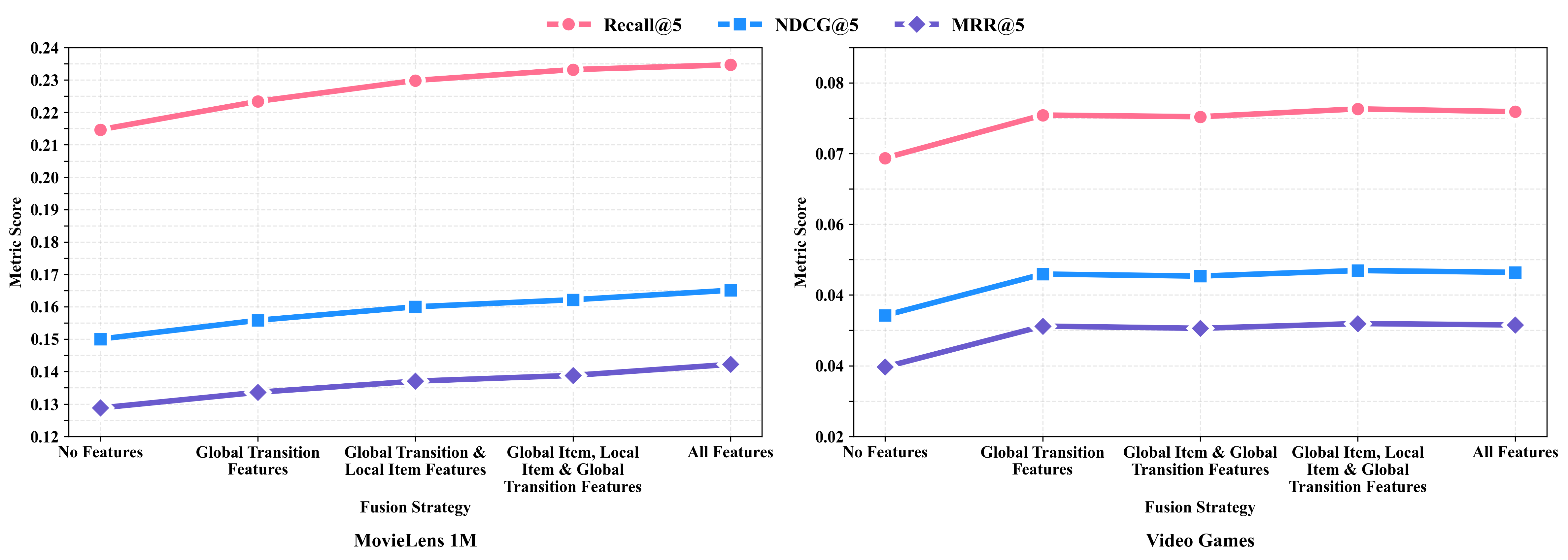}
    \caption{Effect of features on \textit{GrIT} architecture.}
    \label{fig:featureInfluence}
\end{figure}

The results in Figure \ref{fig:featureInfluence} consistently show improved performance with the inclusion of multiple statistical features on the two representative datasets. The variant without group features performs the worst, while the best results are achieved when all four features are jointly utilized for group representation learning. This pattern indicates that each statistical feature captures distinct aspects of user behavior, and their combination is most effective for modeling the multi-faceted nature of user behavior across different temporal scales. Additionally, it also signifies the importance of group representations in modeling the user representation for sequential recommendation.

\subsection{Analysis of Learned Group Representations} 
\label{sec:ablationGroupDist}
To examine the distinctiveness of the learned group profiles, we analyze the similarity structure of the group representations. Specifically, we compute the cosine similarity between group representations ($G^TG$) to assess their diversity. 

\begin{figure}[h]
    \centering
    \begin{subfigure}{0.49\columnwidth}
        \includegraphics[width = \textwidth]{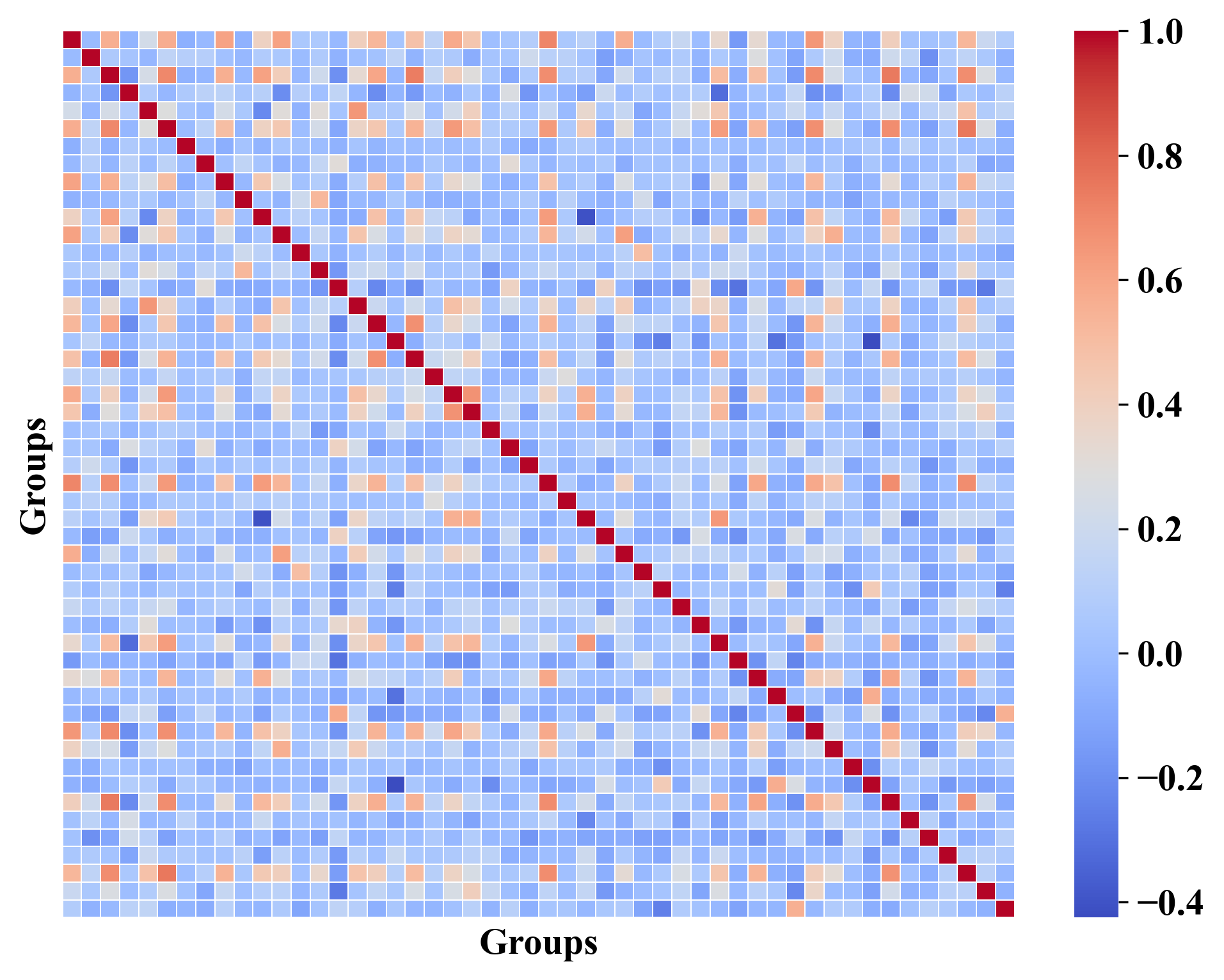}
        \caption{MovieLens 1M}
    \end{subfigure}%
    \begin{subfigure}{0.49\columnwidth}
        \includegraphics[width= \textwidth]{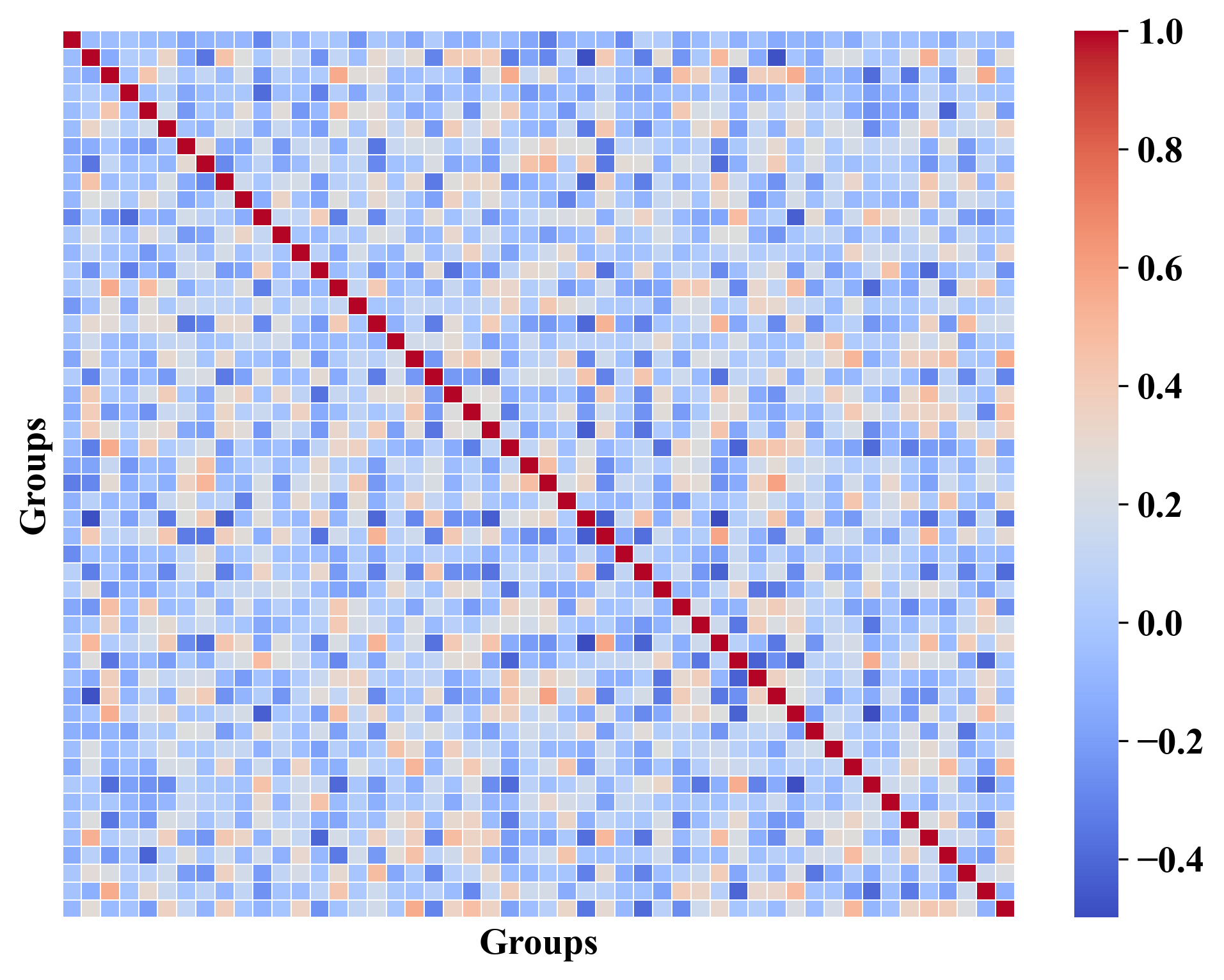}
        \caption{Video Games}
    \end{subfigure}
    \caption{Cosine similarity between groups.}
    \label{fig:groupSim}
\end{figure}

Figure \ref{fig:groupSim} presents the group similarity matrices for two representative datasets. As evident from the figure, the majority of the off-diagonal entries have values close to $0$. This indicates that the learned groups are well-separated and capture distinct behavioral patterns, even without explicit regularization promoting group diversity.

\subsection{Temporal Evolution of Group Profiles} \label{sec:ablationGroupEvolve}

To understand how user preferences evolve over time, we study the temporal dynamics of group memberships learned by \textit{GrIT}. \AS{For this analysis, we randomly select a user and visualize their group membership profile at each timestep. The resulting timeline illustrates how the user's affinity toward different groups evolves with each successive interaction.}

\begin{figure}[h]
    \centering
    \begin{subfigure}{0.49\linewidth}
        \includegraphics[width=\linewidth]{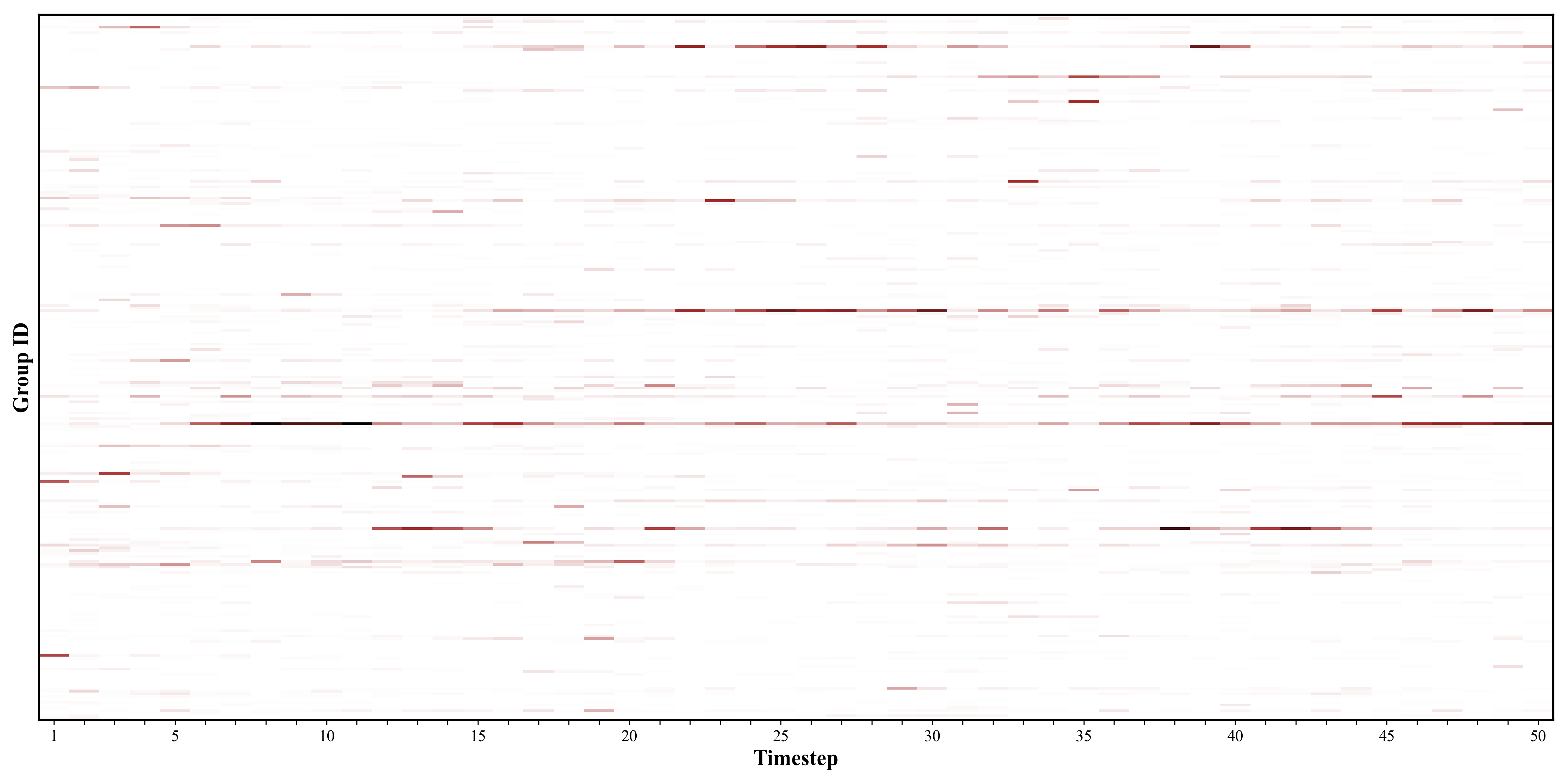}
        \caption{MovieLens 1M}
    \end{subfigure}
    \begin{subfigure}{0.49\linewidth}
        \includegraphics[width=\linewidth]{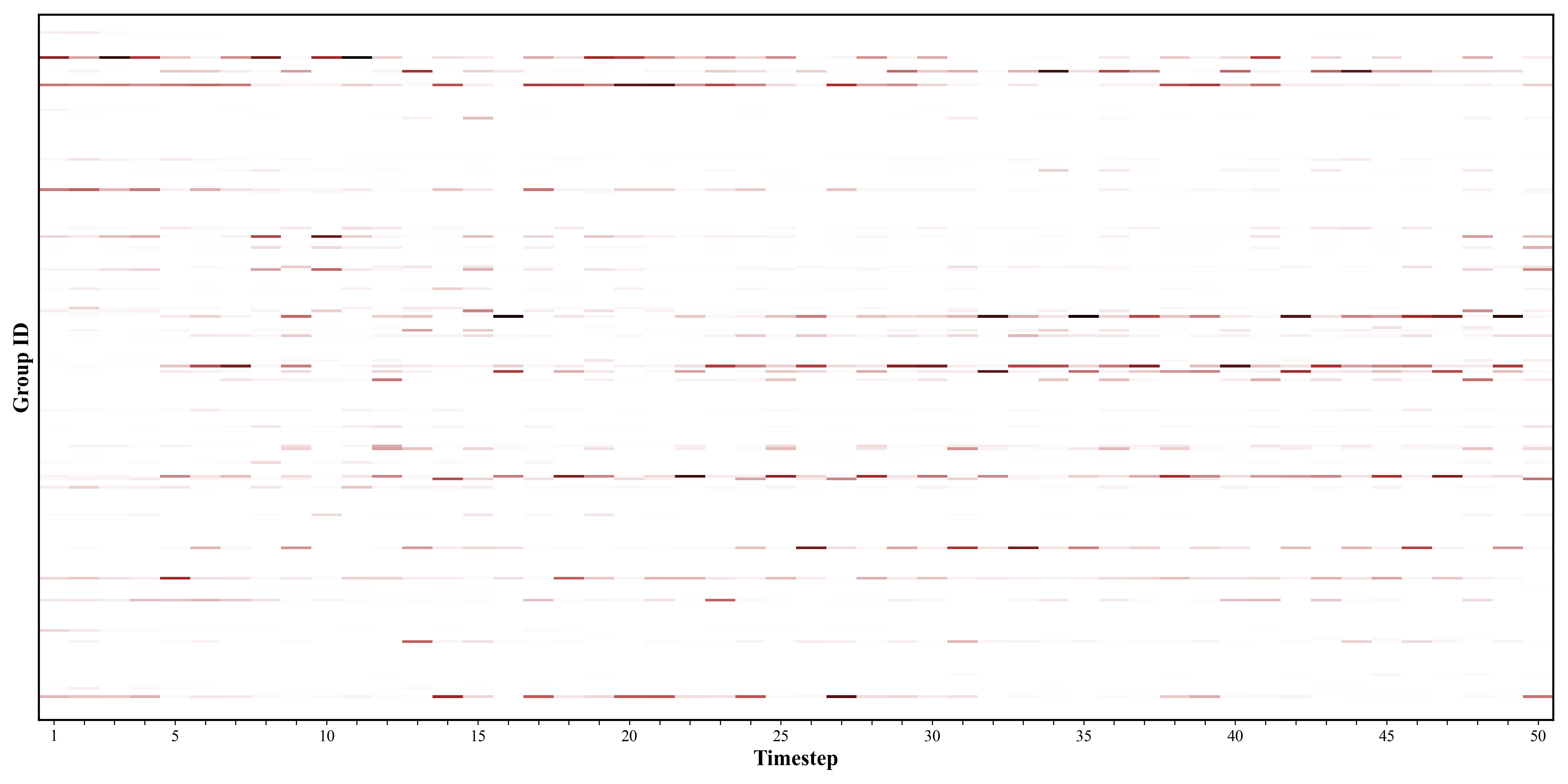}
        \caption{Video Games}
    \end{subfigure}
    \caption{Evolution of group membership over time. \AS{\textit{Darker colors indicate stronger affinity.}}}
    \label{fig:groupEvolve}
\end{figure}

The resulting plots (Figure \ref{fig:groupEvolve}) for two representative datasets reveal that certain groups remain consistently active at most of the timestamps, indicating stable long-term user preferences. Additionally, a few groups appear intermittently, capturing contextual or short-term shifts in behavior. A few isolated activations are also observed, which may correspond to exploratory interests. This analysis confirms the ability of \textit{GrIT} to successfully model the evolving nature of user group memberships, where stable profiles provide continuity while dynamic changes capture adaptation to new preferences.

\subsection{Hyperparameter Sensitivity Analysis of \textit{GrIT}} \label{sec:hyperSensitivity}

We assess the sensitivity of \textit{GrIT} to key hyperparameters, viz., dropout rate, attention dropout rate, the number of groups, $\kappa$, and the control parameter $\beta$. 
\begin{figure}[h]
    \centering
    \begin{subfigure}{0.49\columnwidth}
    \centering
        \includegraphics[width= 0.7\linewidth]{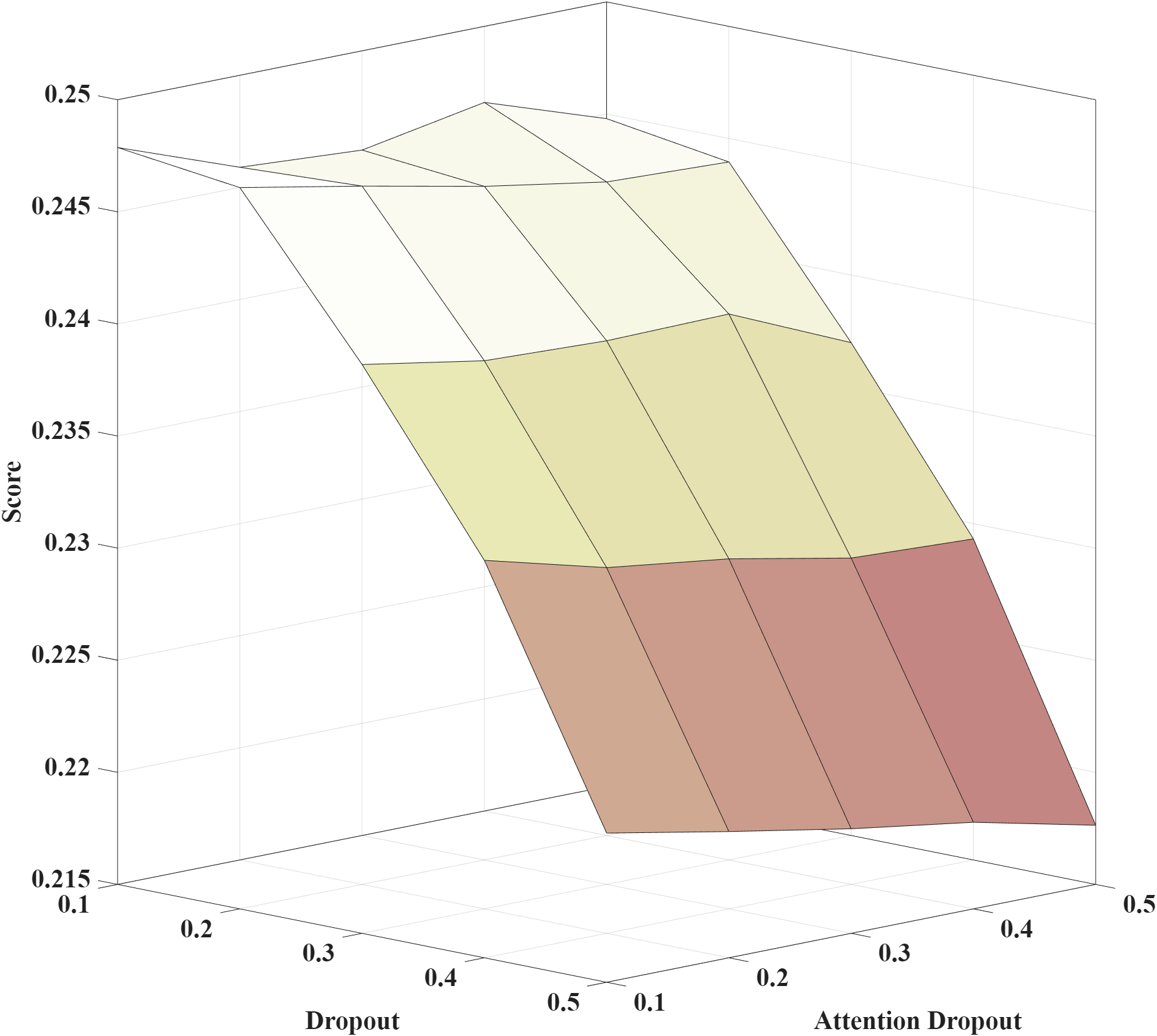}
        \caption{MovieLens 1M}
    \end{subfigure}%
    \begin{subfigure}{0.49\columnwidth}
    \centering
        \includegraphics[width= 0.7\linewidth]{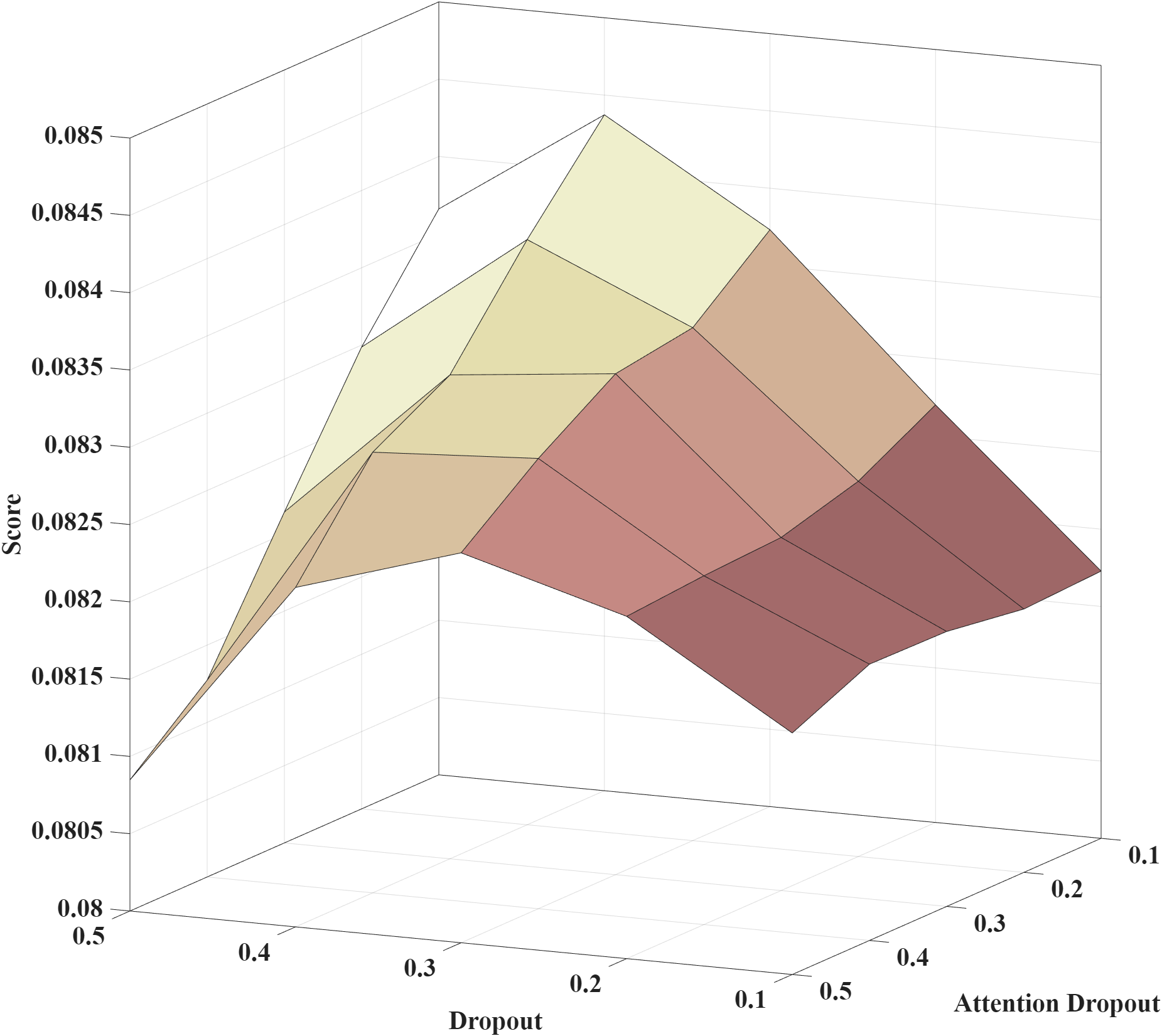}
        \caption{Video Games}
    \end{subfigure}
    \caption{Influence of dropout and attention dropout.}
    \label{fig:dropAttn}
\end{figure}

Figure~\ref{fig:dropAttn} presents a mesh plot of dropout versus attention dropout, showing the average scores of $Recall@10$ and $MRR@10$ on the validation set for two representative datasets. \textit{GrIT} achieves the best performance with lower attention dropout and a higher overall dropout, suggesting that moderate regularization in attention layers, combined with stronger overall dropout, stabilizes training. Similar trends are observed across the remaining datasets.

\begin{figure}[h]
    \centering
    \begin{subfigure}{0.48\columnwidth}
        \includegraphics[width=\linewidth]{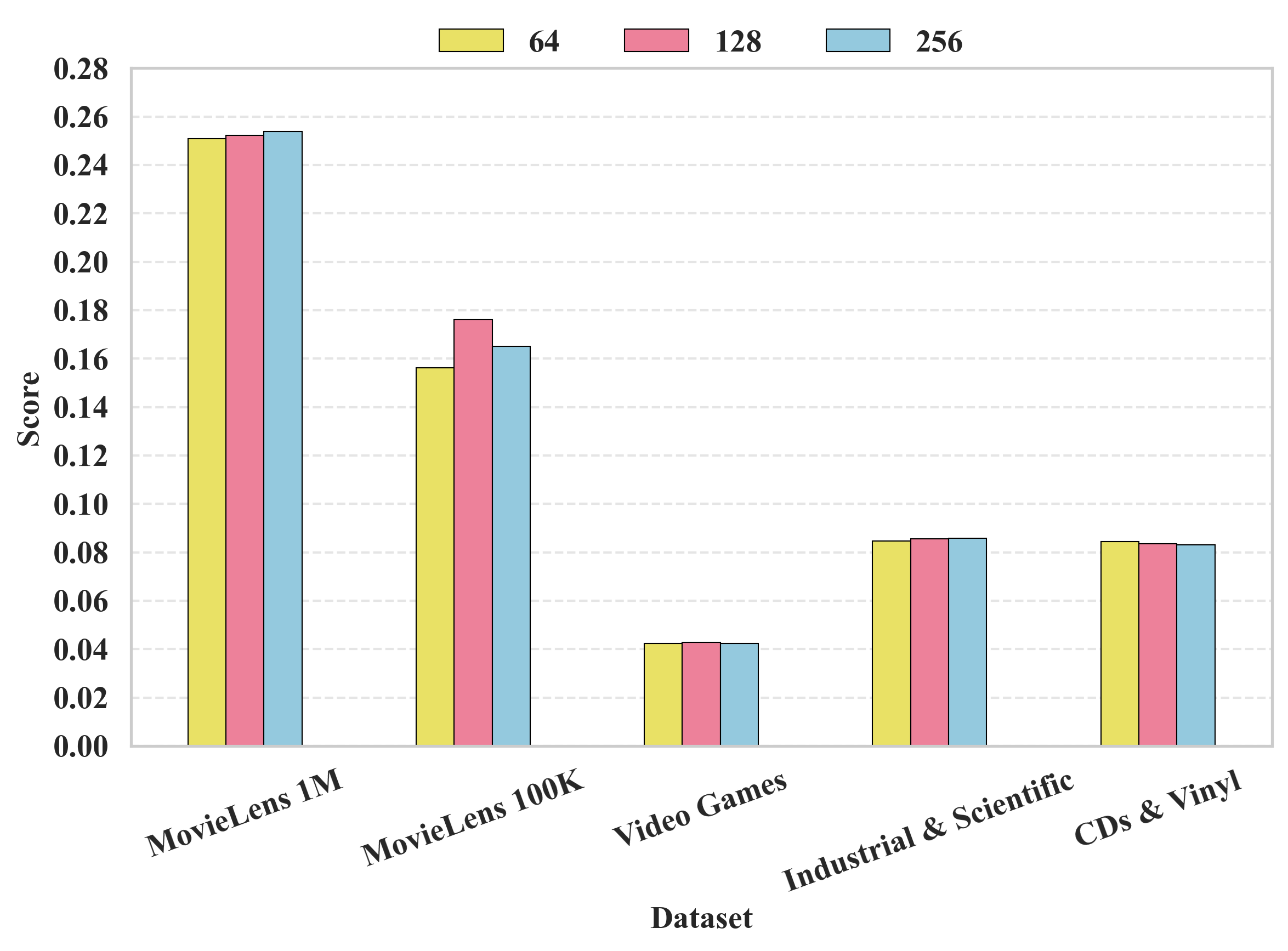}
        \caption{$\kappa$}
        \label{fig:kappa}
    \end{subfigure}
    \begin{subfigure}{0.48\columnwidth}
        \includegraphics[width=\linewidth]{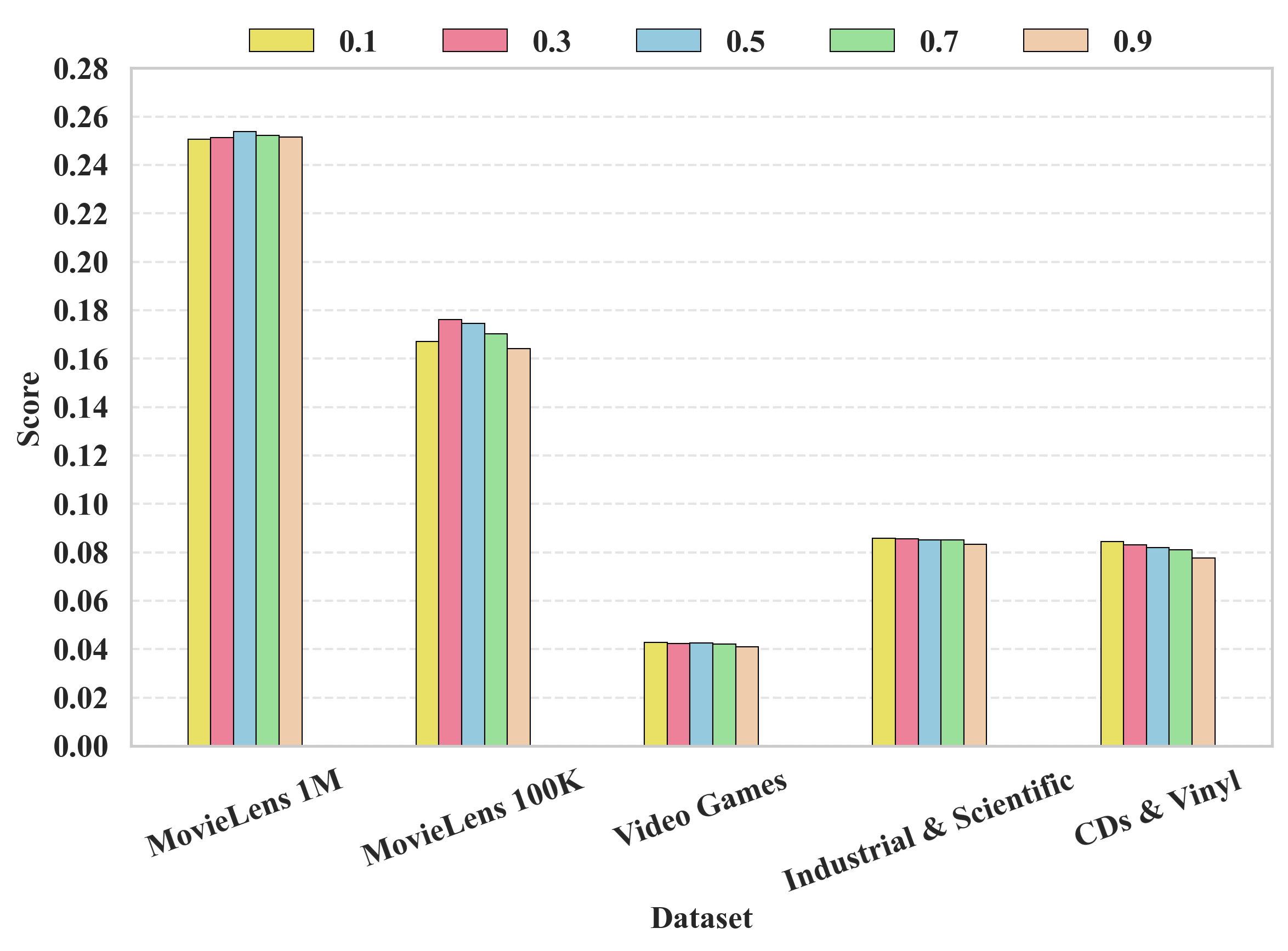}
        \caption{$\beta$}
        \label{fig:beta}
    \end{subfigure}
    \caption{Influence of $\kappa$ and $\beta$.}
    \label{fig:kappaBeta}
\end{figure}

Figure \ref{fig:kappa} and \ref{fig:beta} illustrate the effects of the $\kappa$ and $\beta$ parameters, respectively. For dense datasets, higher $\beta$ values improve performance, whereas sparser datasets benefit from lower $\beta$ values, indicating that group representations are more reliable and informative in richer datasets. Regarding $\kappa$, \textit{GrIT} achieves superior performance with a large number of groups for dense datasets, capturing the diverse behavioral patterns effectively.

\subsection{Statistical Analysis} \label{sec:hypoTest}

To further ensure the validity of the observed results, we conduct a statistical significance analysis. We first apply the \textit{Friedman Test}~\cite{demvsar2006statistical} across all $\mathcal{K}$ competing methods and $\mathcal{N}$ datasets to verify whether there exist statistically significant differences in performance. 
We present results for $Recall@5$, $NDCG@5$, and $MRR@5$ as representative examples, since the performance trends are consistent across other values of $k$.  
The Friedman statistics $F_F$ for each evaluation metric, reported in Table \ref{tab:statisticalAnalysis}, consistently exceed the critical value at $\alpha = 0.05$, thereby rejecting the null hypothesis of equal performance among the models.
\begin{table}[h]
\renewcommand{\arraystretch}{1}
\caption{Summary of the statistical analysis.}
\centering
\adjustbox{max width=\linewidth}{
    \begin{tabular}{lcccccc}
        \hline
        \textbf{Metric} & \textbf{$F_F$} & \textbf{Critical Value ($\alpha=0.05$)} & \textbf{$q_\alpha$} & \textbf{$\mathcal{N}$} & \textbf{$\mathcal{K}$} & \textbf{$CD$} \\ \hline
        {\textbf{Recall@5}} & 21.7353 &  &  &  &  &  \\
        {\textbf{NDCG@5}} & 21.7353 &  &  &  &  &  \\ 
        {\textbf{MRR@5}} & 14.4211 & \multirow{-3}{*}{2.7109} & \multirow{-3}{*}{2.8500} & \multirow{-3}{*}{5} & \multirow{-3}{*}{6} & \multirow{-3}{*}{3.3722} \\ \hline
    \end{tabular}
    }
\label{tab:statisticalAnalysis}
\end{table}

\begin{figure}[h]
    \centering
        \begin{subfigure}[b]{0.33\columnwidth}
            \centering
            \includegraphics[width=\linewidth, trim ={60 63 40 83}, clip]{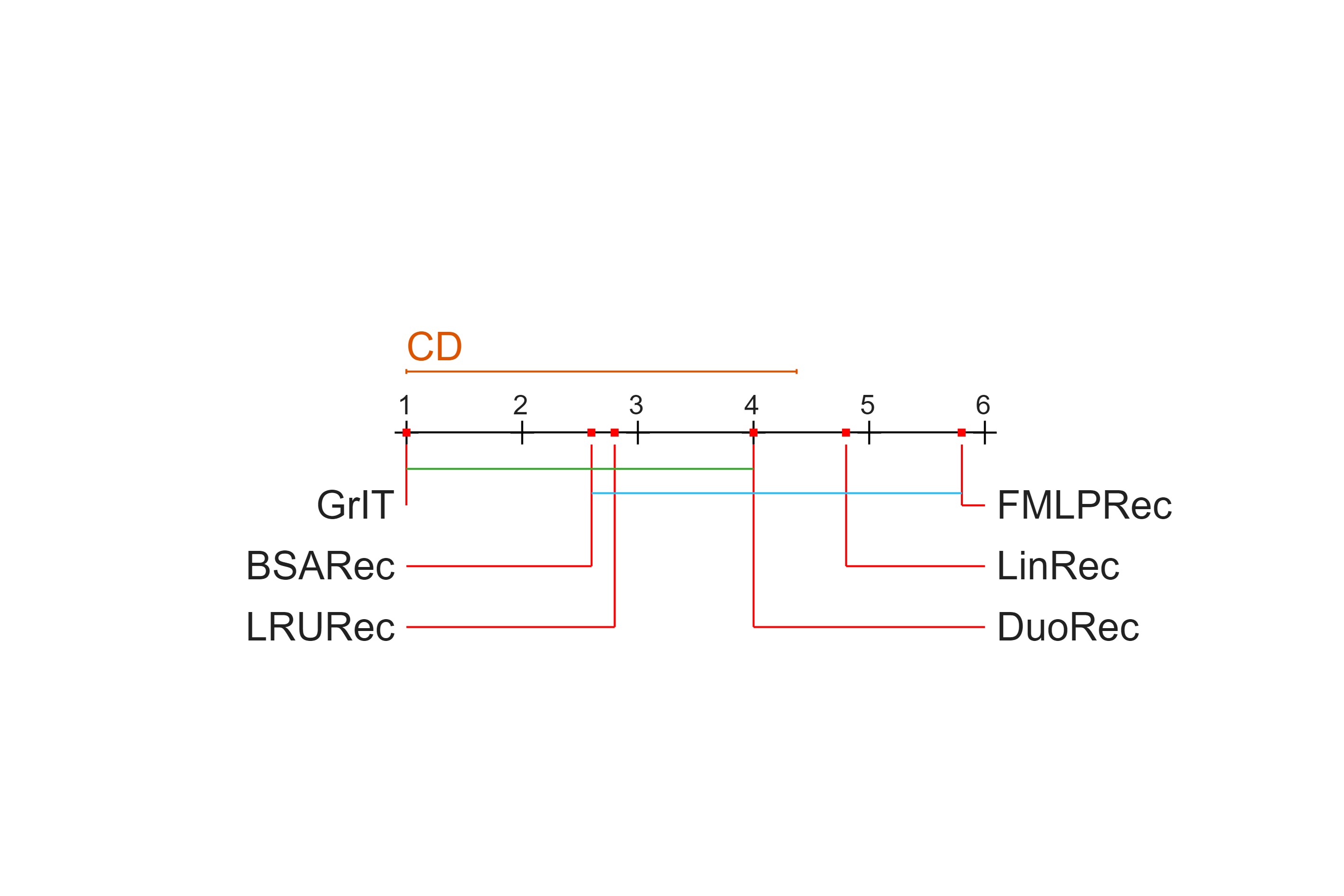}
            \caption{$Recall@5$}
            \label{fig:r@5_nem}
        \end{subfigure}%
        \begin{subfigure}[b]{0.33\columnwidth}
            \centering
            \includegraphics[width=\linewidth, trim ={60 63 40 83}, clip]{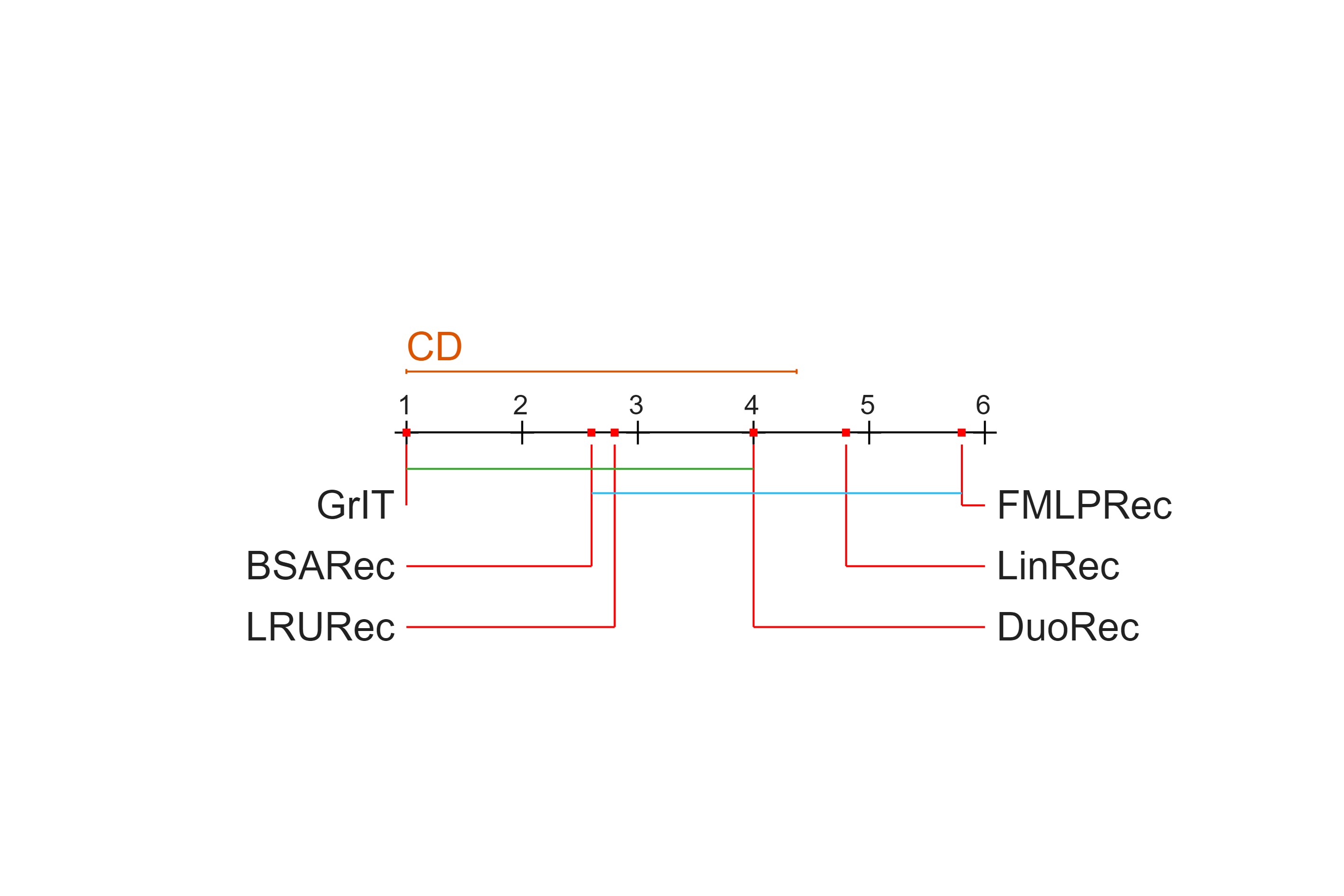}
            \caption{$NDCG@5$}
            \label{fig:n@5_nem}
        \end{subfigure}%
        \begin{subfigure}[b]{0.33\columnwidth}
            \centering
            \includegraphics[width=\linewidth, trim ={60 63 40 83}, clip]{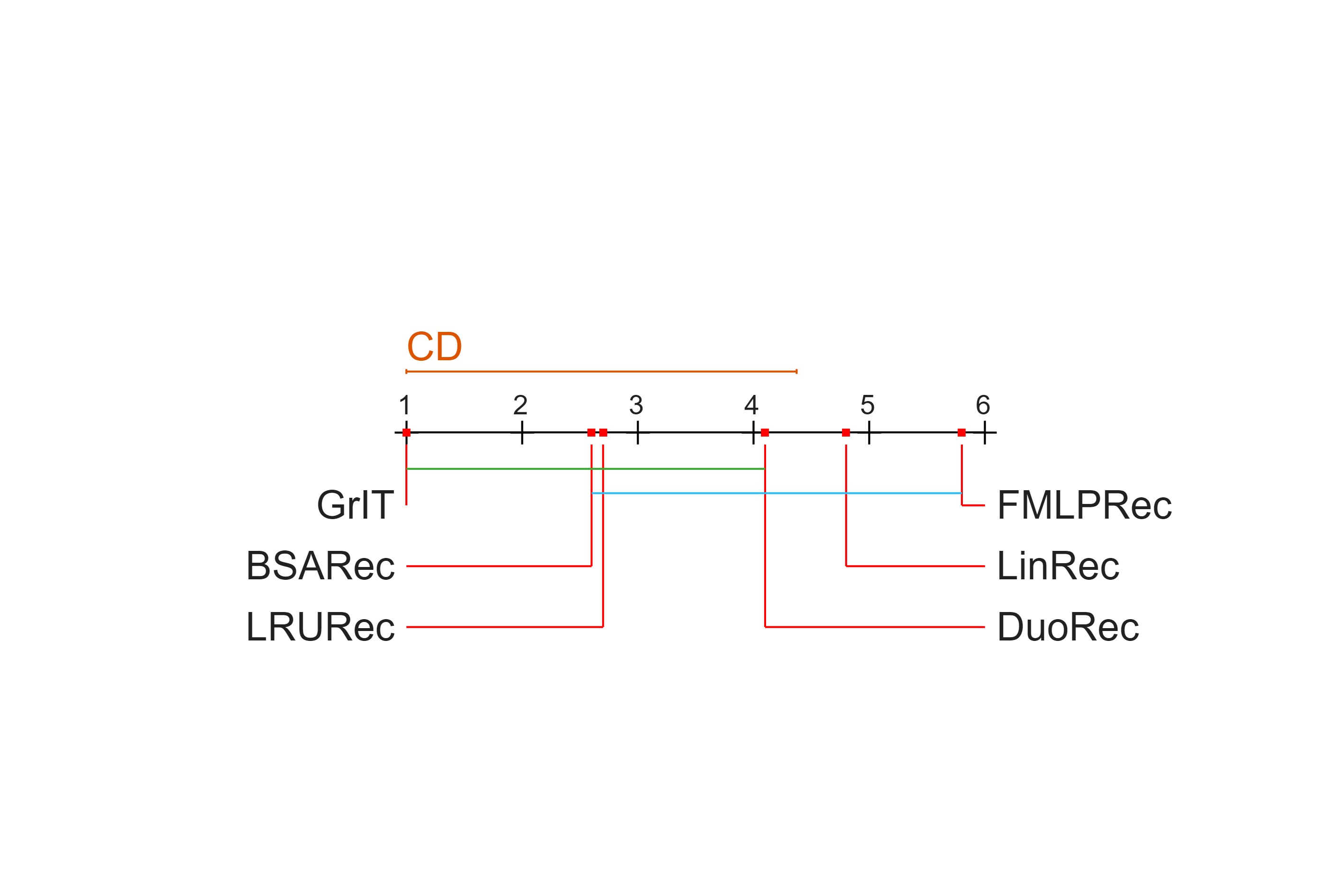}
            \caption{$MRR@5$}
            \label{fig:m@5_nem}
        \end{subfigure}
                
    \caption{CD diagrams for comparing algorithms on $Recall@k$, $NDCG@k$ and $MRR@k$ for $k = 5$.}
    \label{fig:criticalDifference}
\end{figure}

Subsequently, we employ the post-hoc Nemenyi test~\cite{demvsar2006statistical} to identify the pairwise differences. The test computes the critical difference ($CD$) value, beyond which the mean rank difference between two models is statistically significant. Mathematically,
\begin{equation}
    CD = q_{\alpha} \sqrt{\tfrac{\mathcal{K}(\mathcal{K}+1)}{6\mathcal{N}}},
\end{equation} 
where $q_{\alpha}$ denotes the critical value.  The results in Figure \ref{fig:criticalDifference} demonstrate that GrIT maintains a statistically superior rank over the baselines, confirming its reliability and robustness.

\textit{In summary, our experimental study demonstrates the effectiveness of \textit{GrIT}, addressing all seven research questions. The results confirm that \textit{GrIT} not only outperforms state-of-the-art baselines but also benefits from learnable positional encodings and diverse feature combinations for group representation. Our analysis shows that the learned group representations are distinct and well-separated, while user affinities to these groups evolve dynamically over time. Furthermore, GrIT proves to be stable under various hyperparameter settings, and statistical testing validates the significance of its performance improvements. These findings collectively establish GrIT as a robust framework for sequential recommendation.}

\section{Conclusion}
\label{concFuture}

In this work, we propose \textit{GrIT}, a novel group-informed transformer-based model for sequential recommendation that learns user representations by modeling evolving group affinities alongside interaction history
Unlike prior approaches that focus solely on user- and item-centric patterns, \textit{GrIT} learns time-varying group affinity to derive group-based user representation. These group-aware representations are integrated with those obtained from individual interaction sequences, producing context-rich and adaptive user representations for next-item prediction. Comprehensive experiments across five benchmark datasets confirm that \textit{GrIT} consistently outperforms state-of-the-art baselines, highlighting the importance of modeling evolving group dynamics of the user. 
In the future, exploring cross-domain transfer of group embeddings, incorporating multimodal or social signals, and scaling \textit{GrIT} to large-scale, real-time recommendation environments are promising directions for extending \textit{GrIT}. 

\bibliographystyle{unsrt}
\bibliography{References}

\end{document}